\documentclass{article}

\usepackage{arxiv}

\usepackage[T1]{fontenc}

\usepackage{amsmath,epsfig,graphicx,subfigure}
\usepackage{booktabs}
\usepackage{multirow}
\usepackage{url}

\usepackage{graphicx}

\begin{document}

\title{Multimedia Edge Computing} 

\author{Zhi Wang, Wenwu Zhu, Lifeng Sun, Han Hu, Ge Ma, Ming Ma, Haitian Pang, Jiahui Ye, Hongshan Li}

\maketitle

\begin{abstract}

In this paper, we investigate the recent studies on multimedia edge computing, from sensing not only traditional visual/audio data but also individuals' geographical preference and mobility behaviors, to performing distributed machine learning over such data using the joint edge and cloud infrastructure and using evolutional strategies like reinforcement learning and online learning at edge devices to optimize the quality of experience for multimedia services at the last mile proactively. We provide both a retrospective view of recent rapid migration (\emph{resp.} merge) of cloud multimedia to (\emph{resp.} and) edge-aware multimedia and insights on the fundamental guidelines for designing multimedia edge computing strategies that target satisfying the changing demand of quality of experience. By showing the recent research studies and industrial solutions, we also provide future directions towards high-quality multimedia services over edge computing.

\end{abstract}

\maketitle

\section{Introduction}

\subsection{Challenges of Multimedia}

Recent years have witnessed an explosive growth of a variety of types of multimedia data generated at the \emph{edge} of networks, as a result of the rapid development of smartphones, tablets, 5G base stations, mobile routers, and a massive number of devices connected through the Internet of Things (IoT). According to International Data Corporation (IDC) prediction, over $120$ zettabytes of the data will be generated by edge devices by 2025 \cite{zwolenski2014digital}, among which \emph{multimedia} data will be the most significant category. The generation, storage, processing, and delivery of the multimedia data is challenging the conventional centralized data processing and distribution paradigm: it is costly to process them in the centralized infrastructure and brings in the unsatisfactory quality of experience, \emph{e.g.}, large latency and concern of data privacy.

\subsection{Edge Computing for Multimedia}

The new generation of \emph{socialized}, \emph{crowdsourced}, \emph{mobilized}, and \emph{immersive} multimedia services and applications, including mobile social media, interactive livecast, online gaming, and virtual and augmented reality, lead to a large number of new interactions between the participants. The general issues in a centralized solution for the multimedia services above mainly include large edge-to-cloud latency and bandwidth consumption, service availability, energy consumption, and privacy concerns in a centralized solution \cite{shi2016edge}.

These problems have all called for the migration from the conventional centralized cloud multimedia to the distributed \emph{edge multimedia} empowered by the capacities of edge computing and edge network, and/or an mergence between them. The concept of edge computing has been proposed for over a decade, and has been recognized by academia and industry as a trending research area. This paradigm shift is based on the design philosophy to make the infrastructure close to the end-users where the original data is generated, to mitigate the aforementioned latency, bandwidth, energy consumption and privacy issues.

Although the paradigm shift provides a new design space to improve multimedia experiences, with the diverse multimedia data types and requirements in data sensing, edge processing and edge content distribution, a careful study on multimedia edge computing is still in demand.

\subsection{Multimedia Edge Computing}

\begin{figure}[!t]
	\centering
		\includegraphics[width=.8\linewidth]{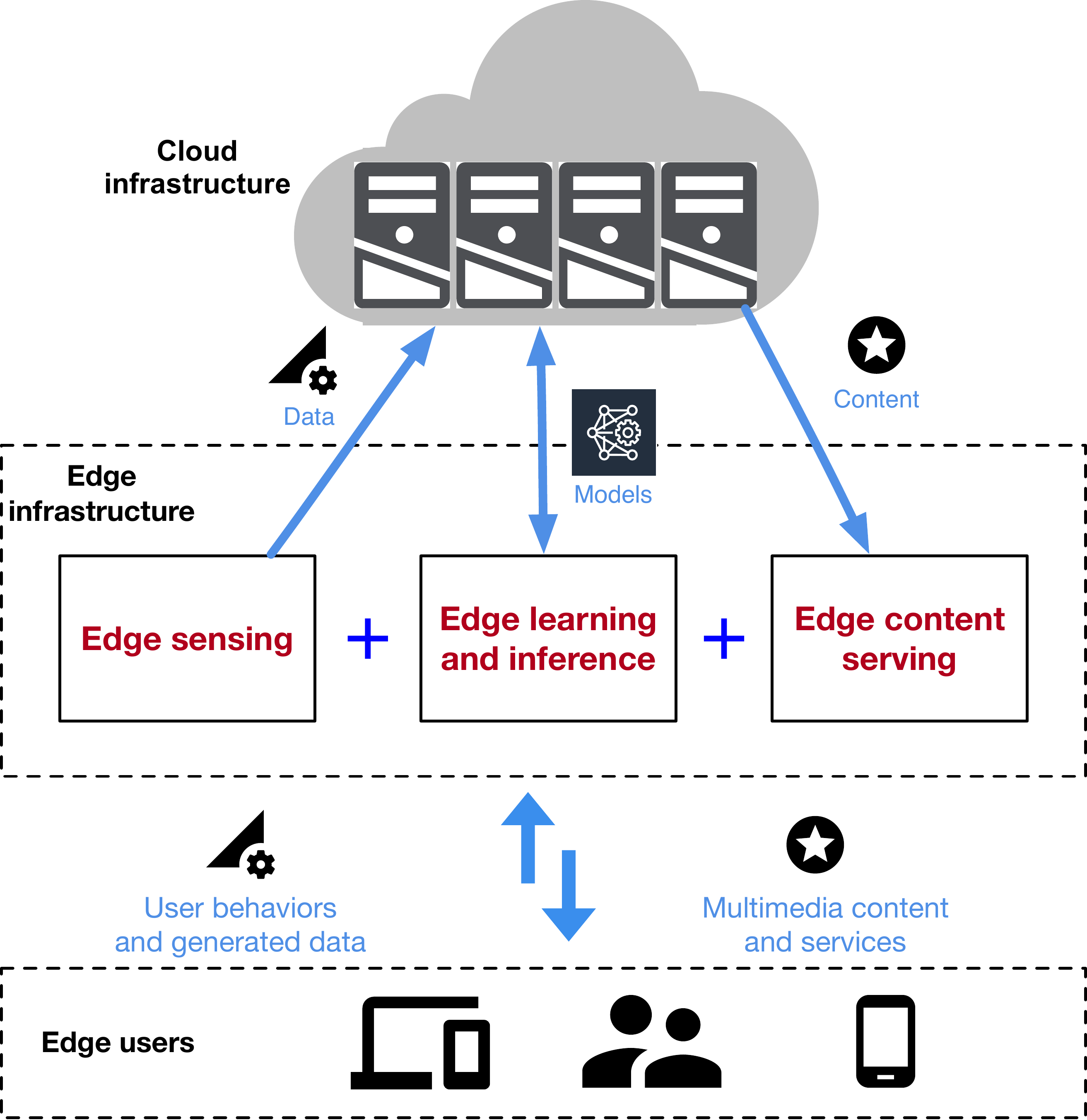}
	\caption{Framework of multimedia edge computing.}
	\label{fig:framework}
\end{figure}

As illustrated in Fig.~\ref{fig:framework}, multimedia edge computing enables a number of multimedia service related procedures to perform at the edge network, including data sensing, synthesizing, machine learning (including both edge training and inference), and content serving (including both caching and distribution). \begin{itemize}

	\item \emph{Edge sensing.} The infrastructure allows users to upload multimedia data to edge devices nearby, which addresses several problems at the ``first-mile'' networks \cite{pang2017first}, where multimedia content is generated in crowdsourced applications, \emph{e.g.}, crowdsourced live streaming like Tiktok. At the same time, the deployment of massive edge devices potentially provide the capabiliteis to sense crowds patterns like geo-preference and mobility, to improve the multimedia service deployment accordingly. 
	
	\item \emph{Edge learning.} With the improvement of computational capacities at edge devices, they are able to perform certain data inference tasks with deep learning models or parts of the models executing on the edge devices \cite{kang2017neurosurgeon, li2018jalad}. Based on the new cloud and edge infrastructure, researchers have started to optimize both the inferece accuracy and infererence latency in a joint framework.
	
	\item \emph{Edge serving.} Using edge infrastructure to cache content that is to be requested by users, can significantly reduce the backbone load and end-to-end latency; while for mechanism design, in edge content caching and delivery, not only content popularity but also the mobility patterns of users affect these strategies \cite{ma2017understanding}. Meanwhile, due to the increased dynamics in multimedia content request patterns caused by the ``edgeness,'' reinforcement learning and online learning frameworks have been investigated to improve the adaptation capabilities for such solutions \cite{garg2019online}.
	
\end{itemize}

Most of the existing survey studies usually investigate edge computing for general purposes. Mao \emph{et al.}~\cite{mao2017mobile} provided a research outlook from the centralized mobile cloud computing towards mobile edge computing. Abbas \emph{et al.}~\cite{abbas2017mobile} studied the importance and challenges of the deployment of mobile edge computing, and the impact of edge computing integration with the traditional mobile and cloud networks. From the architecture perspective, Mach \emph{et al.}~\cite{mach2017mobile} studied computation offloading using the edge infrastructure, including decision on computation offloading, allocation of computing resource within the mobile edge computing and mobility management. Roman \emph{et al.}~\cite{roman2018mobile} have analyzed the security threats, challenges, and mechanisms inherent in all edge paradigms.

In this paper, we study edge computing from a \emph{multimedia} perspective, covering edge content generation, processing, and consumption. In particular, we present the general paradigms of multimedia edge computing and review studies on multimedia edge sensing, edge inference and edge content distribution.

The rest of the paper is organized as follows. In Section \ref{sec:paradigm} we present the paradigms including multimedia-aware edge computing and edge-aware multimedia. In Section \ref{sec:sensing} we investigate multimedia data sensing from the edge network. In Section \ref{sec:learning} we show how the joint edge-cloud infrastructure is able to provide scalable and low-latency edge inference for multimedia services. In Section \ref{sec:caching}, we study learning-based edge caching strategies for the dynamical changing multimedia requests at edge networks. Finally, we conclude the paper and provide some potential future directions in Section \ref{sec:conclusion}.

\section{Paradigms of Multimedia Edge Computing} \label{sec:paradigm}

\subsection{Multimedia and Edge Computing}

Based on the relationship between multimedia and edge computing, the concept of multimedia edge computing can be classified into two categories: \emph{multimedia-aware edge computing} and \emph{edge-aware multimedia}. \begin{itemize}

\item In multimedia-aware edge computing, previous studies are devoted to designing better edge-computing infrastructure and strategies, with the awareness of multimedia services and applications, to handle multimedia tasks and processing multimedia content. In our study, both edge sensing and edge content serving belong to this category, in a sense that edge servers and resources are allocated to improve the traditional multimedia content serving. 

\item In edge-aware multimedia, studies are focusing on making multimedia data collection, storage, and processing, more ready to take advantage of the availability of the existing edge computing. In our study, the edge learning and inference belong to this category, since deep learning models are ``changed,'' \emph{e.g.}, by model distillation or compression, to make use of the existing edge resources. \end{itemize}

\subsection{Services of Multimedia Edge Computing}

Based on the serving stage, we also divide multimedia edge computing into edge sensing, edge learning and edge content serving. In particular, we are going to explore the potentials of multimedia edge computing, in improving the capabilities of content sensing and harvesting, machine learning and content caching. \begin{itemize}

\item For edge sensing, we reveal how edge-sensed data can be utilized for improving the multimedia service quality. We consider two respresentative studies. The first one is data-driven edge behavior sensing, where user behaviors are sensed from edge network. The second one is crowdsourced multimedia edge \emph{harvesting}, which explores improving the performance for users to upload the user-generated multimedia content with edge-network path \emph{re-routing} at the first mile, to enhance the experinece in crowdsourced content generation. For example, the trending live streaming has been optimized to utilize edge devices as the initial streaming relays, so as to improve the broadcaster's upload quality, which in turn improves the overall streaming quality for viewers significantly \cite{pang2018content}.

\item For edge learning, we present that the combination of edge computing and machine learning provides a promising solution with the purpose of enhancing the quality and speed of inference over deep learning models. Edge inference processes and analyses the data or a certain fraction of the personal data locally, which effectively protects users' privacy, reduces response time, and saves bandwidth resources. Being aware that specific inference tasks (\emph{e.g.}, detection for different objects) are deployed on different edge devices, machine learning techniques including transfer learning \cite{pan2009survey}, federated learning \cite{konevcny2016federated} and meta learning \cite{finn2017model} have also been applied in multimedia tasks with new inference requirements, such as low latency, heterogeneous sample distributions, etc.

\item For edge content caching and delivery, user requests can be served with much smaller latency if the edge caches are deployed close to the users. An edge caching policy can thus be either reactive or proactive: a reactive caching policy determines whether to cache a particular content after it has been requested, while a proactive caching strategy places content at edge nodes actically before users request it, based on the prediction of content popularity. A good example is mobile social multimedia content distribution. Based on the mobility sensing capability enabled by edge networks, crowds' mobility patterns are inferable from their associations with base stations nearby, and a content provider can thus proactively deploy content likely to be requested by users in the neighborhood to reduce the latency when they actually fetch them \cite{ma2017understanding}. \end{itemize}

Next, we will present the recent research studies and industrial solutions for these different multimedia edge computing services. 

\section{Sensing from the Edge} \label{sec:sensing}

In this section, we present how edge-sensed data can be utilized for improving the multimedia service quality. In particular, by jointly investigating crowd mobility patterns and their preferences, new intelligence can be developed \cite{ma2017understanding}.

\subsection{Data-driven Edge Behavior Sensing}

The first case study is to understand user behaviors in edge networks. 

\subsubsection{User Preference}

One representative application with edge sensing is mobile video streaming. In \cite{ma2017understanding}, we studied mobile video behaviors in the edge networks. Datasets used for this study covers how users view videos in the mobile video streaming app has been recorded in $2$ weeks. In each trace item, the following information is recorded: 1) The device identifier, which is unique for different devices and can be used to track users; 2) The timestamp when the user starts to watch the video; 3) The location where the user views the video: the video player reports the location either collected from the device's built-in GPS function or inferred from the network parameters (\emph{e.g.}, cellular base station); 4) The title of the video.

\subsubsection{Edge-sensed Mobility}

The data for studying user mobility patterns covers over $1$ million Wi-Fi APs in Beijing city, including the basic service set identifier (BSSID) of Wi-Fi APs and the location of the Wi-Fi hotspots. This dataset samples a large fraction of Wi-Fi APs that are actually deployed in Beijing, allowing us to determine whether these APs can provide content delivery functionality for mobile video streaming. Each trace item contains the latitude and longitude of the AP and the point of interest (PoI) information of the AP (\emph{e.g.}, hotel). The dataset also contains cellular network information, including locations, IDs, and location area code (LAC) of the cellular base stations.

\subsection{Temporal and Spatial Request Patterns} \label{sec:mobile-video}

To study the mobility patterns of viewers, we assume that the users' requests can be served by the nearest Wi-Fi APs or cellular BSes. Thus, we first classify all the users in the mobile video streaming system into two categories: \emph{multi-location users}, who request videos in different locations (APs/BSes) within \emph{one day} in the traces, and \emph{single-location users}, whose requests are all issued from the same place (APs/BSes) within \emph{one day}. Note that a user may be a multi-location user or a single-location user on different days.

\subsubsection{Skewed Geographical Request Distribution}

We investigate the geographical distribution of requests. According to the longitude interval 0.01$^{\circ}$ and latitude interval 0.01$^{\circ}$, the region of the whole city is divided into different locations. Every location can be regarded as a 0.01$^{\circ}$ $\times$ 0.01$^{\circ}$ geographic location with an area of $0.72 km^2$. Each location has a PoI functionality label, which indicates the largest PoI functionality number of the location. We count the number of requests issued in these locations. Our observations are as follows: 1) More requests are issued at night than during the daytime, \emph{e.g.}, the number of requests from 6 pm--12 pm is $74\%$ greater than that from 12 am--6 pm. 2) A significant fraction of locations only has very few requests issued. These observations indicate that to serve mobile video requests, the edge network content delivery systems need to take the geographical request distribution into consideration, \emph{e.g.}, to allocate more resources to the locations with higher request density and proactively push content to the edge at the off-peak times.

\subsubsection{Video Requests on the Move}

We study the behaviors of multi-location users in different locations within one day, such as a university, airport, railway station, scenery spot, and business district. Our observations are as follows: 1) These locations generally have a relatively stable multi-location user fraction of approximately $20\%$. 2) Some locations have lower multi-location user fractions than others, \emph{e.g.}, there are fewer users in university than at the rail station. 3) \emph{The fraction of multi-location users changes significantly over time in some locations}, \emph{e.g.}, the fraction in the business distinct drops from approximately $25\%$ on weekdays to $15\%$ on weekends. The reason is that mobile video behaviors are highly correlated with the regular commute behaviors of users.

\subsubsection{Periodical Request Patterns}

To specify the periodical request patterns, we use a frequency analysis approach \cite{wang2015understanding}, as follows: \begin{enumerate}

	\item[i)] Let $\mathbf{x} = (x_1,x_2,\ldots,x_N)^{T}$ denote the number of video requests over time, \emph{i.e.}, $x_i$ is the number of requests in time slot $i$. In our experiments, each time slot is $1$ hour, and we study the request samples in $1$ week, \emph{i.e.}, $N=168$.
	
	\item[ii)] We perform DFT as follows,
	$$
		X[k] = \sum_{n=1}^{N}x_n e^{-2\pi i kn/N}, \label{eq:dft}
	$$
	where $X[k]$ is the frequency spectrum of the sequence of requests $X$ in the time domain. A larger $X[k]$ indicates that the sequence has a more substantial period of $k$.
	
	\item[iii)] We study the amplitude of the frequency-domain sequence $X[k]$, in which amplitude and phase represent request volume and their peak-valley time, respectively.

\end{enumerate}

\begin{figure}[!t]
     \begin{minipage}[t]{.48\linewidth}
          \centering
                \includegraphics[width=\linewidth]{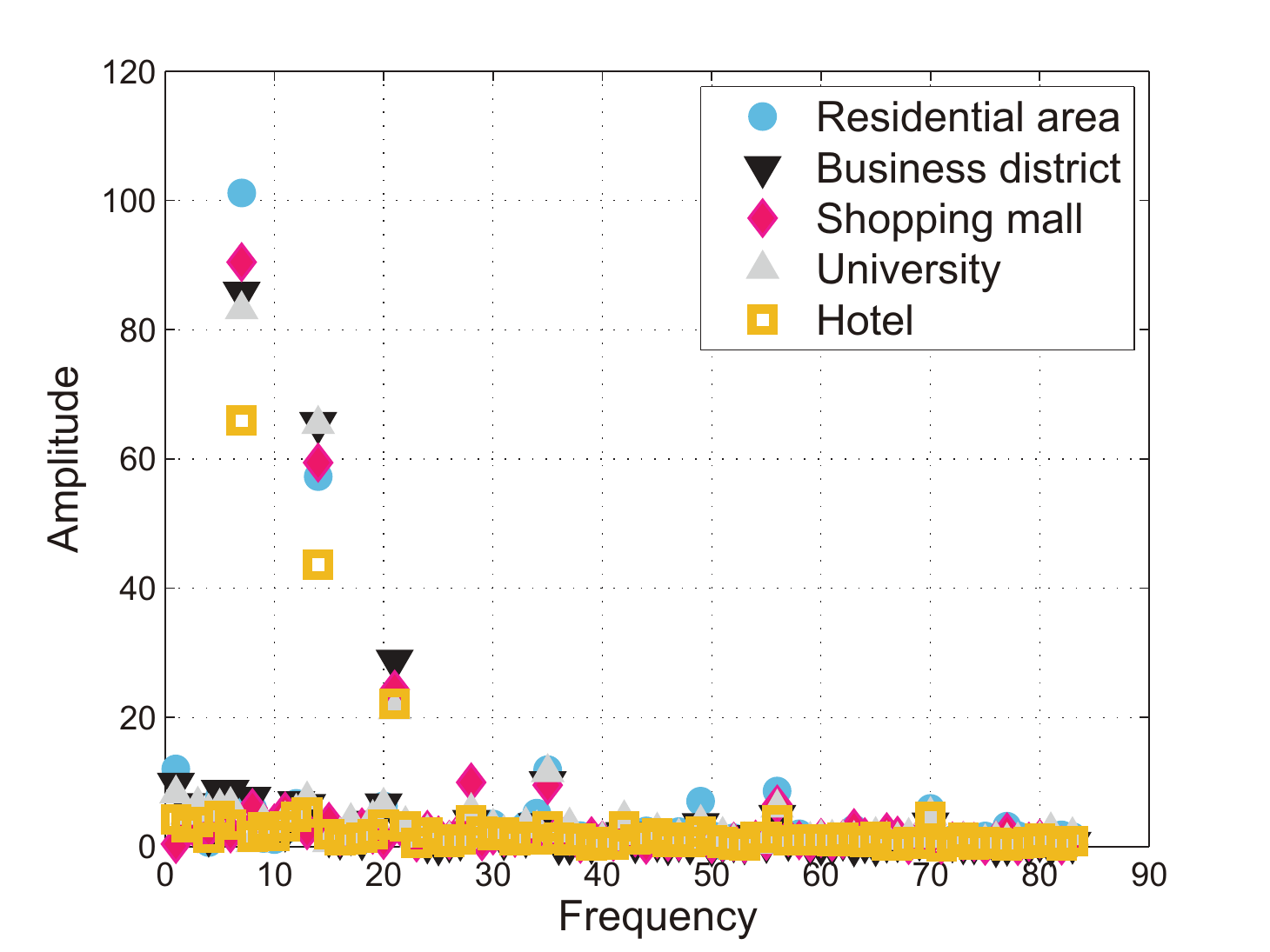}
          \caption{Frequency analysis of mobile video requests in different locations.}
          \label{fig:frequency}
     \end{minipage}
     \hfill
     \begin{minipage}[t]{.48\linewidth}
          \centering
               \includegraphics[width=0.7\linewidth,angle = 90]{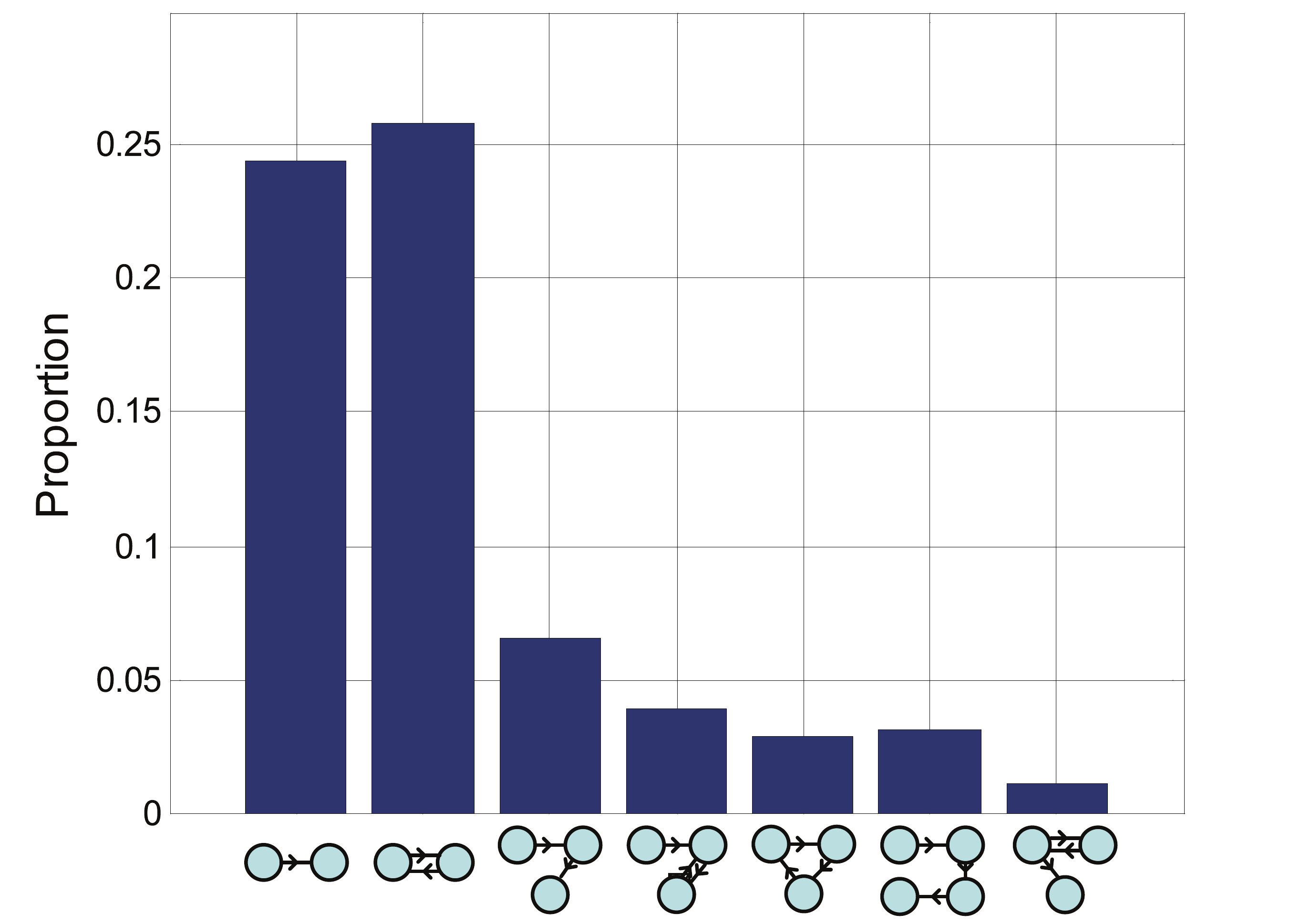}
          \caption{Fraction of migration patterns.}
          \label{fig:mig-pattern}
     \end{minipage}
\end{figure}

Fig.~\ref{fig:frequency} shows the discrete Fourier transform (DFT) results of the requests over time. In particular, we plot the amplitude versus the frequency of requests in different functionalities of locations. Our observations are as follows: 1) There are some major frequencies with large amplitudes, \emph{e.g.}, $k = 7, 14,$, and $21$, corresponding to $1$ day, $12$ hours and $8$ hours, respectively. This means that we can use the three frequency components to present the time-domain traffic. Furthermore, we can leverage this property to predict future traffic. 2) Different functionalities of locations also have different major frequency patterns. For example, the daily pattern is more evident for the residential areas than the hotels, and the business areas have a strong period of $8$ hours. This observation indicates that \emph{the periodical patterns of mobile video requests are highly affected by the functional type of locations}, which can be utilized to distinguish locations with different functionalities.

\subsection{Video Requests Affected by Mobility Behaviors} \label{sec:userpatterns}

In this section, we study what drives the previous request patterns. Particularly, we focus on mobile video user behaviors. In the following experiments on multi-location users, our results are the average results of fourteen days.

\subsubsection{Mobility Intensity Analysis}

In our experiments, we only study the behaviors of \emph{active users} who requested at least ten videos daily in our 2-week traces. Among these $9,576$ \emph{active users}, we have $30\%$ \emph{multi-location users} and $70\%$ \emph{simple-location users}, which are defined previously.

\textbf{Locations Visited}

We first study the mobility intensity of the multi-location users. In Fig.~\ref{figure{fig:locations:a}}, we plot the fraction of users versus the number of ``movements,'' \emph{i.e.}, the number of requests issued in different locations in \emph{one day}. We observe that the number of movements is generally in the range $[1,30]$, and the range $[2,3]$ has the largest fraction of users. The results are quite similar for weekdays and weekends. We next study the number of locations where the requests are issued. In Fig.~\ref{figure{fig:locations:b}}, the bars are the fraction of users versus the number of locations where videos are requested in one day. As shown in this figure, as many as $50\%$ of the multi-location users only issued video requests at $2$ locations, and $80\%$ of the users only requested videos from less than $4$ locations. These results indicate that \emph{it is common for multi-location users to request videos from different locations, but the number of locations (per user) is quite limited}. It provides some basic characteristics to capture the trajectory of multi-location users.

\begin{figure}[!t]
	\centering
		\subfigure[Movement number]{
			\label{figure{fig:locations:a}}
			\includegraphics[width=0.47\linewidth]{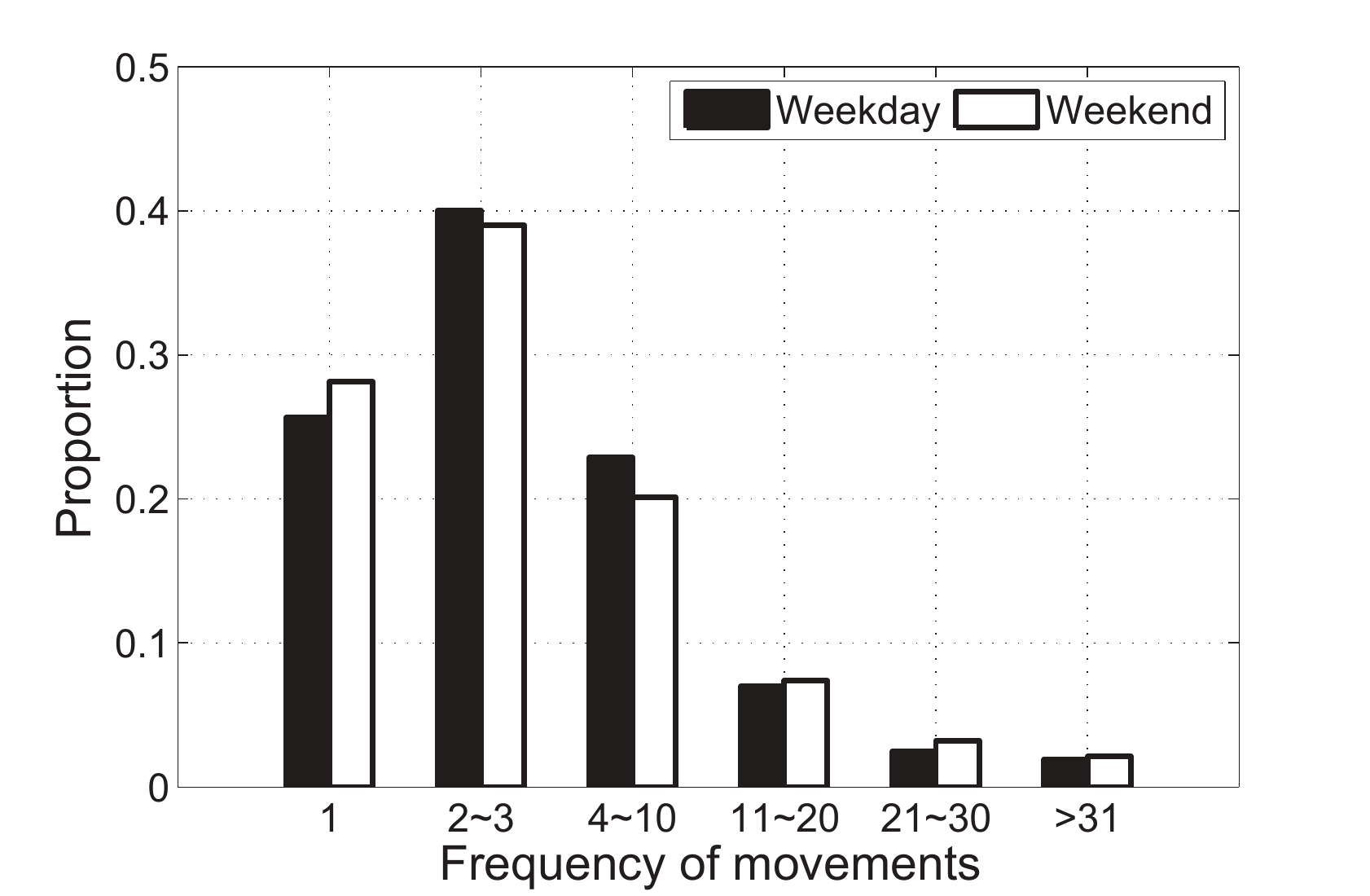}}
		\hfill
		\subfigure[Location number]{
			\label{figure{fig:locations:b}}
			\includegraphics[width=0.47\linewidth]{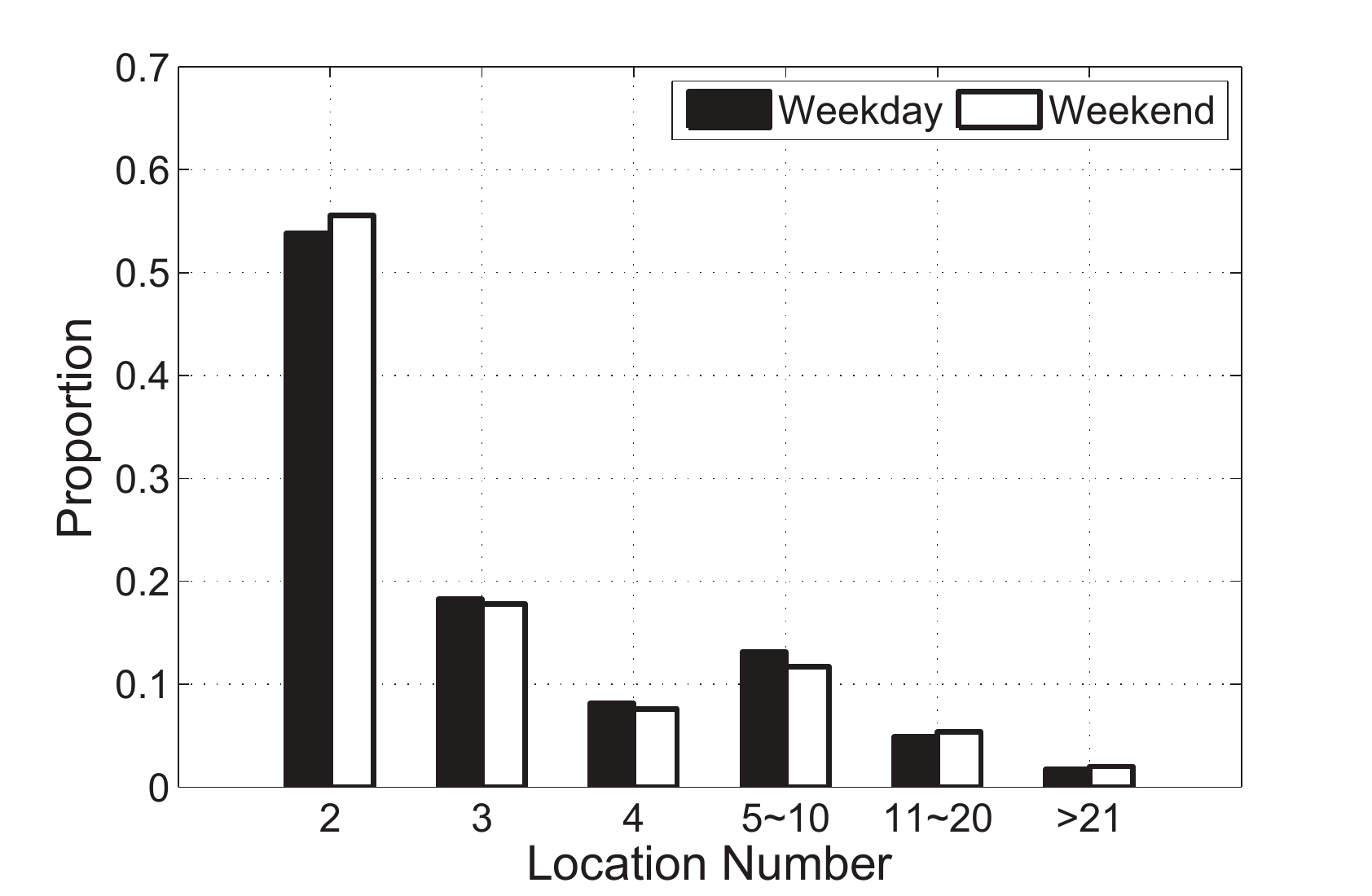}}
	\caption{Statistics of movement of mobile users.}
	\label{figure{fig:locations1}}
\end{figure}

\textbf{Distance and Interval of Movements}

\begin{figure}[!t]
	\centering
		\subfigure[Distance]{
			\label{fig:distance}
			\includegraphics[width=0.47\linewidth]{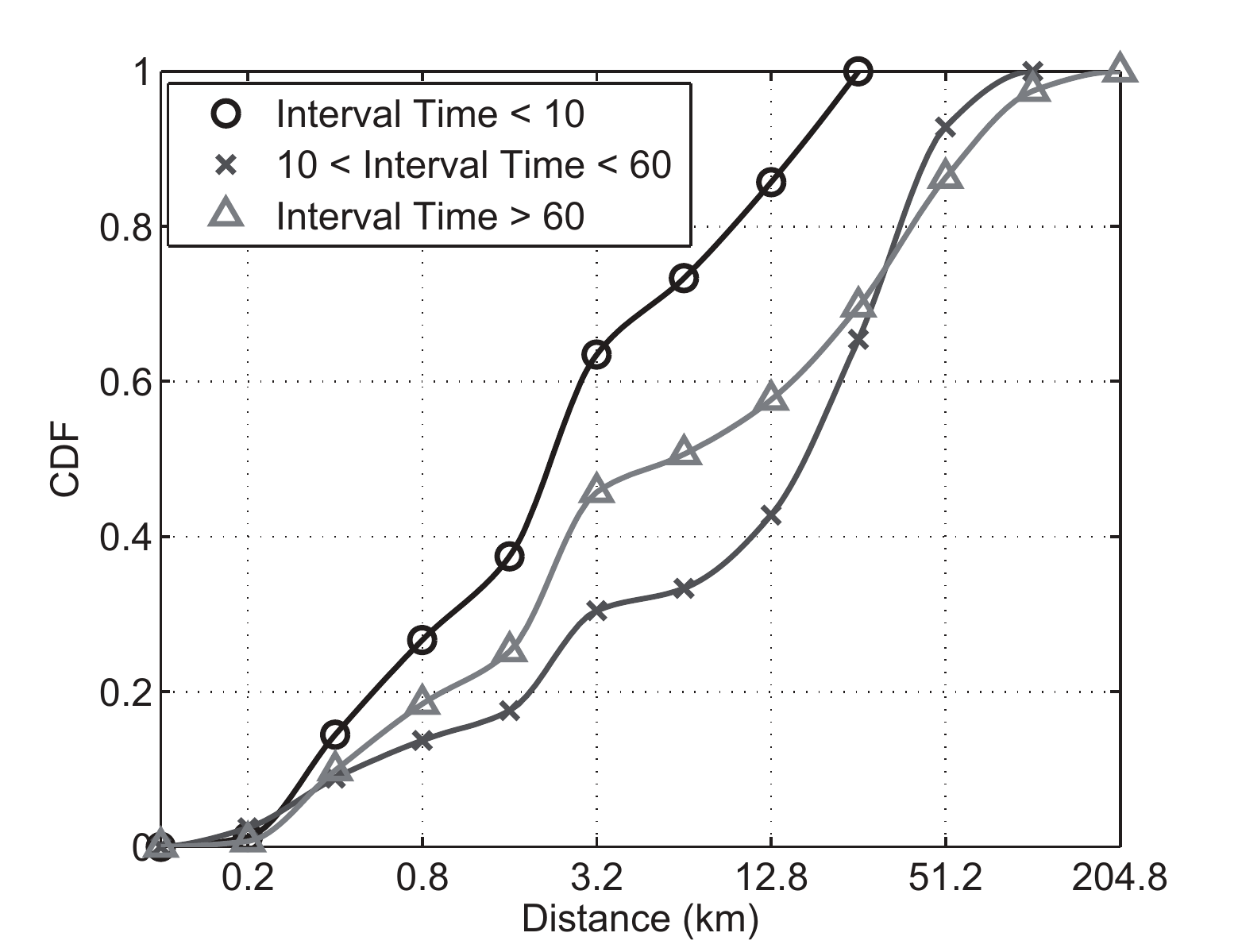}}
		\hfill
		\subfigure[Interval]{
			\label{fig:interval}
			\includegraphics[width=0.47\linewidth]{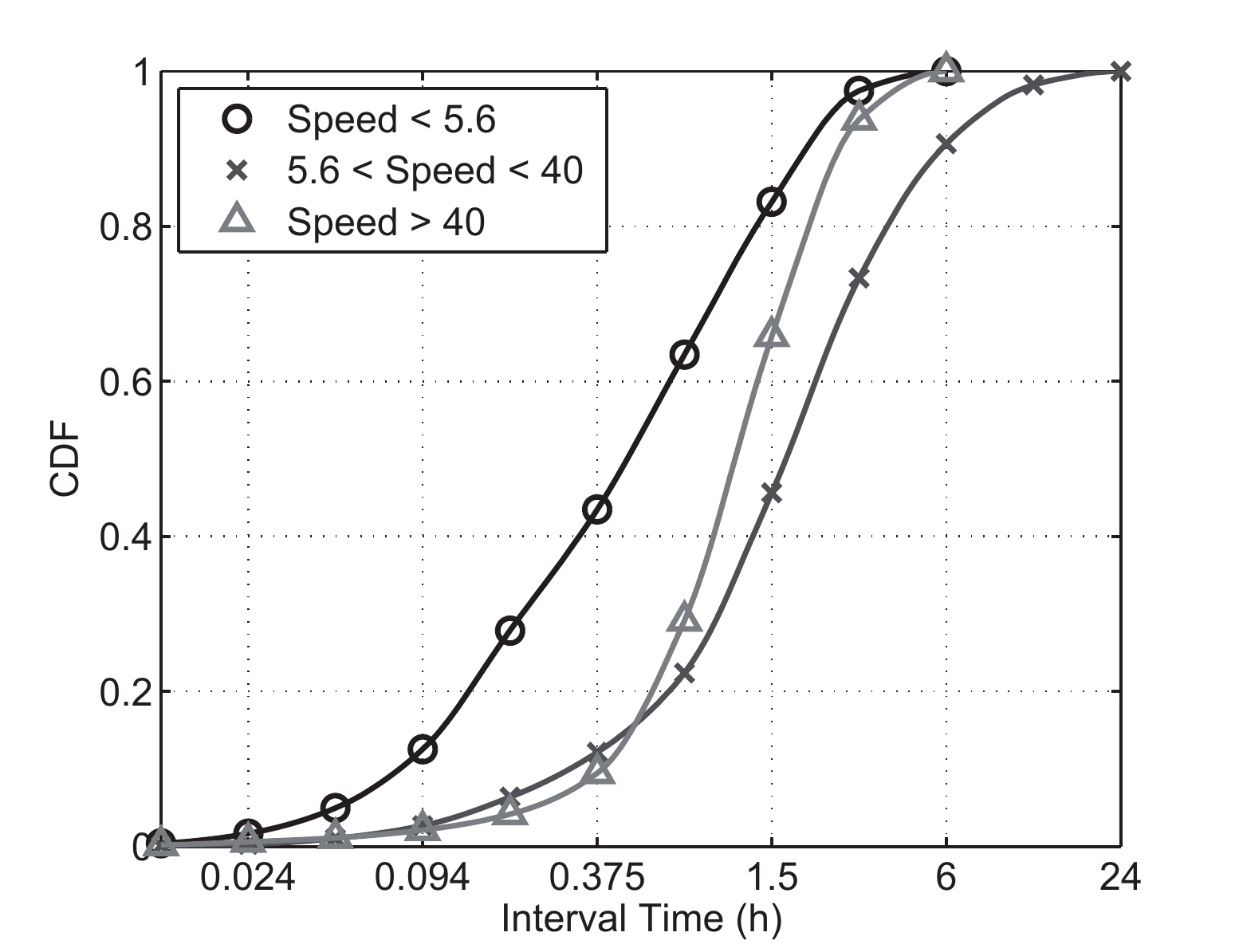}}
	\caption{CDFs of distances and intervals of consecutive requests.}
	\label{figure{fig:locations}}
\end{figure}

We further measure the cumulative distribution of the distances between consecutively visited locations with different time intervals. Fig.~\ref{fig:distance} plots the CDFs of distances between locations where users consecutively request mobile videos. In detail, we select $3$ intervals to divide users into the same order: $[0,10)$ min, $[10,60)$ min, and $[60,\infty]$ min. We observe that when the interval is shorter than $10$ min, the distance is much shorter than that with the other intervals. However, as the interval time increases, the distance does not always become longer. The small time interval indicates that users frequently move between different locations. It is inferred that most users move between $2$ or $3$ locations in a small time interval.

We also study the intervals between consecutive mobile video requests. In Fig.~\ref{fig:interval}, we plot the interval between successive requests of users with different movement speeds. We choose two reference speeds: the average speed of walking (\emph{i.e.}, $5.6$ km/h) and the average speed of the subway (\emph{i.e.}, $40$ km/h). As shown in this figure, when the speed is less than $5.6$ km/h, most of the request intervals are small. For example, $80\%$ of the request intervals are issued within $1.5$ hours, whereas only $40\%$ of the request intervals are issued in $1.5$ hours for the movement speed of $[5.6,40)$ km/h. These observations indicate that the mobility speed of users also affects the request patterns. Moreover, these results largely depend on vehicles, which determine the route time.

\subsubsection{Migration Patterns}

For the multi-location users who request videos in different locations, we study their migration patterns, \emph{i.e.}, how they move across different locations.

According to our previous observations, users only request mobile videos in a small number of locations. We study how they move across these locations. In Fig.~\ref{fig:mig-pattern}, we plot the fraction of users who share the same migration patterns across different locations. In this figure, we plot the most popular $7$ migration patterns, which contribute $70\%$ of all the migrations between locations. We observe that \emph{moving between two particular locations constitutes almost $50\%$ of the migrations}. Additionally, there are migration patterns across $3$ and $4$ locations. These results provide us with the essential characteristics to construct connections between different locations for achieving caching cooperation strategies.

\subsection{Crowdsourced Multimedia Edge Harvesting}

Besides sensing the preference of users and their mobility patterns, edge computing and edge network also enable us to explore improving \emph{content harvesting} with \emph{path re-routing} at the first-mile.

\subsubsection{Impact of First-Mile Network}

Different from conventional on-demand multimedia services (\emph{e.g.}, Hulu or Netflix) and professional live streaming services (\emph{e.g.}, ESPN), a new type of crowd broadcast services faces an especially severe challenge for low latency and sustainable bandwidth as the most broadcasters employ ordinary mobile devices and unstable network for live streaming. A major reason that makes edge harvesting a better solution is as follows: the first-mile connection between broadcast and the server is usually a bottleneck. 

Zhang \emph{et al.}~\cite{zhang2015crowdsourced} showed that in live broadcast services, a viewer rebuffer might either be caused by the first mile at the broadcaster's uplink or the last mile at the viewer's downlink. If a network slowdown (\emph{e.g.}, caused by congestion) is happening at the first mile, the streaming server fails to receive the video chunks, causing every viewer to encounter live interruption, which appears to be rebuffering on his/her video player. First-mile network degradation causes severe quality issues. Pang \emph{et al.}~\cite{pang2018content} used a metric of \emph{viewer rebuffer percentage} (\emph{i.e.}, the number of rebuffered viewers divided by the number of all viewers in the same channel) to study the impact of first-mile network quality, and observed that the situation that $100\%$ of the viewers are encountering rebuffer, accounts for $17\%$ of all rebuffer events; that is, as large as $17\%$ of the rebuffer events are likely to be caused by the first-mile network quality.

\subsubsection{Edge Routing to Improve Upload Quality}

Pang \emph{et al.}~\cite{pang2018content} introduced an edge harvesting network, to enable data re-routing at the application layer, to reduce the network sudden slowdown probability. In this framework, some ``relays'' are provided by an edge network operator. When a ``relayed path'' outperforms the default path, one can choose it to reduce latency and obtain sustainable bandwidth \cite{savage1999end}. The relay assignment problem is based on the network performance, and keeping track of the network performance between the relay and server is feasible. With the network performance collection from direct measurement and prediction, the network performance keeps up to date, which can be used for relay assignment. A hybrid assignment solution is proposed to determine the relay assignment at a different time and broadcaster number scales. The hybrid solution determines whether a broadcast streaming should be relayed, and further which edge relay to use. The centralized assignment takes the whole end-to-end network performance information as input, to calculate the optimal relay assignment of all broadcasters as output. Due to relatively large computations, the centralized assignment operates periodically.

When a broadcast channel is established, it should be relayed to an optimal path at once. The centralized assignment is not appropriate in this scenario, as the centralized assignment cannot guarantee a fast response. Hence, a distributed assignment has been designed to make a quick decision for a better network condition in subsecond response time when a broadcast channel is established. The relay decision is performed on the broadcasters’ devices and relay nodes in a distributed way. Specifically, the broadcaster’s device decides which relay to use with multi-armed bandits \cite{vermorel2005multi}, which explores the network conditions based on the historical data. Once a broadcast channel is assigned to a relay, the relay nodes then decide which server to upload based on the current traffic information.

To summarize, edge sensing provides the capability to perform data-driven investigation on request patterns and usres' trajectories, which can be utilized to guide fine-grained edge multimedia serving to be presented later. Meanwhile, edge itself serves the first-mile data harvesting infrastructure: multimedia content can be uploaded better with strategies like edge re-routing.

\section{Scalable and Low-latency Edge Inference} \label{sec:learning}

Most of today's deep learning models are deployed on dedicated devices in the central datacenter. In this scheme, end users have to upload a large amount of input data (\emph{e.g.}, images and video clips) to the servers, causing high transmission latency. To address this problem, Chen \emph{et al.}~\cite{chen2015glimpse} presented an object tracking system that drops video frames from the raw video to improve bandwidth efficiency. Jain \emph{et al.}~\cite{jain2015overlay} suggested using the blurred frames to reduce the upload traffic load. The inherent limitations of these conventional cloud-only studies are that they have to upload the original data to the cloud, causing high network traffic load and transmission latency. A clear trend is that edge and cloud infrastructures are utilized more and more jointly to provide low latency, high accuracy, and efficient execution in deep multimedia inferencing.

\subsection{Latency Issue of Centralized Inference}

With the great success of deep learning in computer vision, this decade has witnessed an explosion of deep learning-based computer vision-based applications. Because of the enormous computational resource consumption for deep learning applications (\emph{e.g.}, inferring an image on VGG19~\cite{VGG19} requires $20$ GFLOPs of GPU resource), in today's online computer vision-based applications, users usually have to upload the input images to the central cloud service providers (\emph{e.g.}, SenseTime, Baidu Vision and Google Vision, etc.), leading to a significant upload traffic load. 

Several existing studies have been following the \emph{edge-offloading} scheme. Ran \emph{et al.}~proposed a framework that deciding running a deep neural network model either on the edge side or the cloud side based on a function fitted on battery consumption, latency, and current network condition \cite{ran2018deepdecision}. Jiang \emph{et al.}~proposed to use edge devices to extract global information that is unavailable for a single client, to assist the deployment of deep learning applications \cite{jiang2018chameleon}, so they can reduce the redundant computation of each stream.

\subsection{Challenges of Using Edge Inference}

\subsubsection{Lack of information about the cloud-based computer vision models} 

Previous studies~\cite{DeepN-JPEG, torfason2018towards, gueguen2018faster}, generally assume that the details of the computer vision models are available so that they can adjust the JPEG configuration according to the model structure, \emph{e.g.}, one can train a model to determine the JPEG configuration by plugging the original computer vision model into it. However, the structural details of online computer vision models are usually proprietary and not open to the users. 

\subsubsection{Different cloud-based computer vision models need different JPEG configurations} 

As an adaptive JPEG configuration solution, we target to provide a solution that is adaptive to different online computer vision-based services, \emph{i.e.}, it can \emph{generate} JPEG configuration for different models. However, today's cloud-based computer vision algorithms, based on deep and convolutional computations, are quite hard to understand. The same compression quality level could lead to a different accuracy performance. This phenomenon is  presented in \cite{delac2005effects} and commonly seen in adversarial neural network researches~\cite{yuan2019adversarial, evtimov2018robust}.

\subsubsection{Lack of well-labeled training data} 

In edge inference, a problem is that one is not provided the well-labeled data on which image should be compressed at which quality level, as in conventional supervised deep learning tasks. In practice, such an image compression module is usually utilized in an online manner, and the solution has to learn from the images it uploads automatically.

\subsection{Joint Accuracy- and Latency-aware Model Decoupling}

To deploy the deep learning application using the computational power on edge, Li \emph{et al.}~proposed a joint accuracy- and latency-aware deep learning model decoupling solution \cite{li2018jalad}. 

\begin{figure*}[!t]
	\centering
		\includegraphics[width=\linewidth]{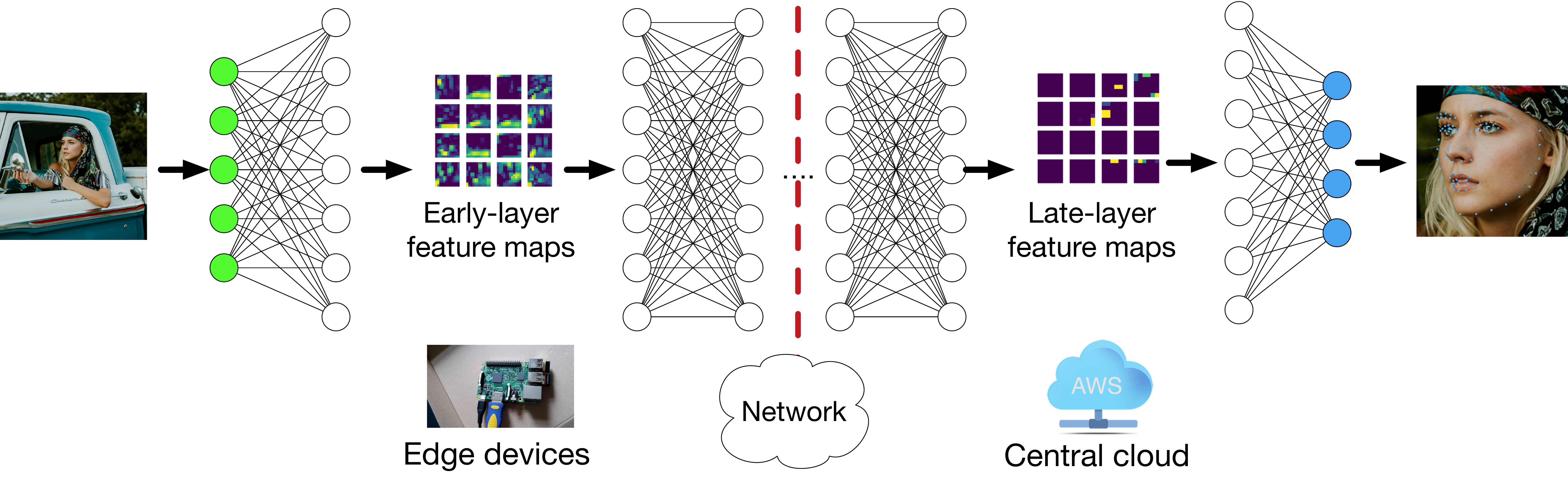}
	\caption{Framework of the joint accuracy- and latency-aware deep network decoupling.}
	\label{fig:decoupling-framework}
\end{figure*}

The framework of the joint accuracy- and latency-aware deep network decoupling design is presented in Fig.~\ref{fig:decoupling-framework}. The deep learning models share a layered network structure, making it possible to decouple a deep learning model layer-wise and deploy it separately. Therefore, the models are decoupled as follows: 1) Feed the raw data (\emph{e.g.}, an image) to the input layer of the deep learning model and run some of its first layers on the edge device; 2) Compress the feature map of the last layer on the edge device, and transmit it to the cloud server, in which runs the late layers of the deep learning model; 3) The decoupling point layer will dynamically change according to the user accuracy demand, network condition and the computation capacity on the edge device.

\subsubsection{Decoupling points}

Not all connection points between layers can be used for decoupling. In traditional deep neural networks like AlexNet, VGGNet, input data flow straightforward along with layers. However, some recent deep learning models tend to introduce more complex network structures rather than sequential models (\emph{e.g.}, ResNet, GoogleNet and InceptionNet ) \cite{Res_Net, GoogleNet, InceptionNet}. To be general to all deep learning models, logical decoupling points are defined, such that a full \emph{res-unit} in ResNet is regarded as one hyper layer, equivalent to a convolution layer in conventional sequential models. It means that the granularity is layer-wise when decoupling a sequential model, while the granularity is unit-wise when decoupling a complex branchy structure.

\subsubsection{Accuracy-aware In-layer Feature map Compression}

Motivated by the fact that the in-layer feature maps have strong sparsity, the in-layer feature maps can be compressed to reduce the in-layer data size while trying to keep the accuracy loss within a user-defined boundary. The in-layer feature map compression is composed of two steps.

First, almost all current deep learning models use float-point values to express the feature values, which takes up a considerable data size, the float feature values are converted into low-bit integers, quantize the float values into integers by using a step conversion function shown as below. Such quantization has been proved feasible to compress the parameters.

Next, after the quantization, a large number of values are mapped to  $0$. Therefore one can conduct further compression with benefit from conventional variable-length code. Huffman Coding is then used to compress the sparse integer feature maps further.

\subsection{Reinforcement Learning-based Edge Compression}

Besides model decoupling, with the assistance of multimedia edge infrastructure, one can also compress the data at edge before uploading it for inference. Though JPEG has been used as the \emph{de facto} image compression and encapsulation method, its performance for the deep computer vision models is not satisfactory. Liu \emph{et al.}~\cite{DeepN-JPEG} showed that by re-designing the quantization table in the default JPEG configuration, one can compress an image to a smaller version while maintaining the comparable inference accuracy for a deep computer vision model.

Li \emph{et al.}~\cite{li2019adacompress} present a reinforcement learning-based solution, called AdaCompress, to choose a proper compression quality level for an image to a computer vision model on the cloud-end, in an online manner. First, they have designed an interactive training environment that can be applied to different online computer vision-based services. A Deep Q-learning Network-based~ \cite{DQN} agent is proposed to evaluate and predict the performance of a compression quality level on an input image. In real-world application scenarios, this agent can be highly efficient to run on today's edge infrastructures (\emph{e.g.}, Google edge TPU).
	
A reinforcement learning-based framework is proposed to train the agent in the above environment. The agent can learn to choose a proper compression quality level for an input image after iteratively interacting with the environment by feeding the carefully designed reward that considers both accuracy and data size. An \emph{explore-exploit} mechanism is designed to let the agent switch between ``sceneries.'' In particular, after deploying the agent, an \emph{inference-estimation-querying-retraining} solution is designed to switch the RL agent intelligently once the scenery changes and the existing running agent cannot guarantee the original accuracy performance.

\begin{figure}[!t]
	\begin{minipage}{.9\linewidth}
		\centerline{\includegraphics[width=\linewidth]{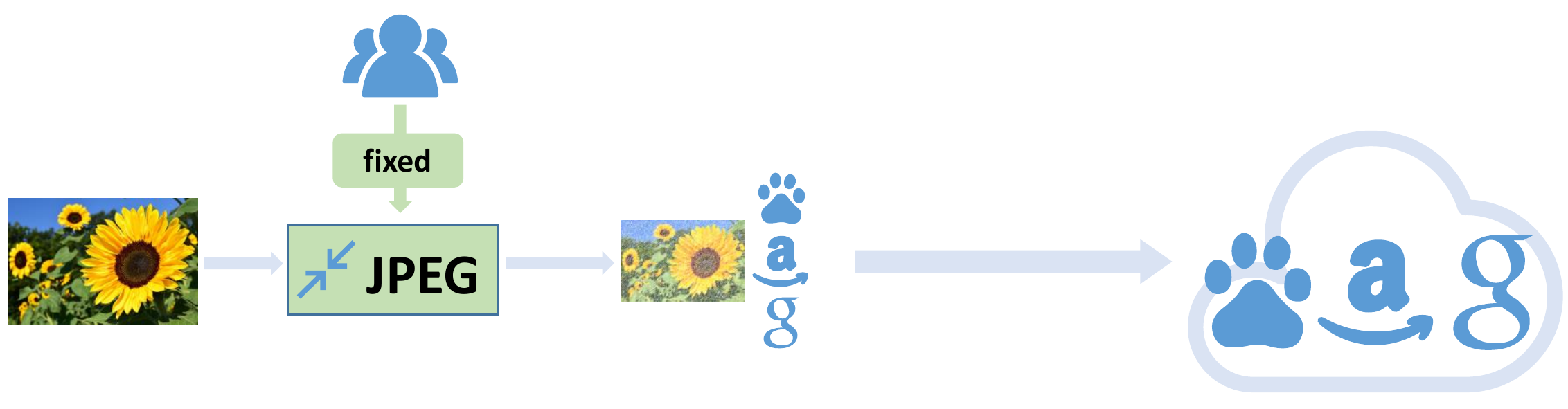}}
		\begin{center}
			{(a) Conventional solution: fixed manually defined compression level}
		\end{center}
		\vspace{0.3cm}
	\end{minipage}	
	\begin{minipage}{.9\linewidth}
		\centerline{\includegraphics[width=\linewidth]{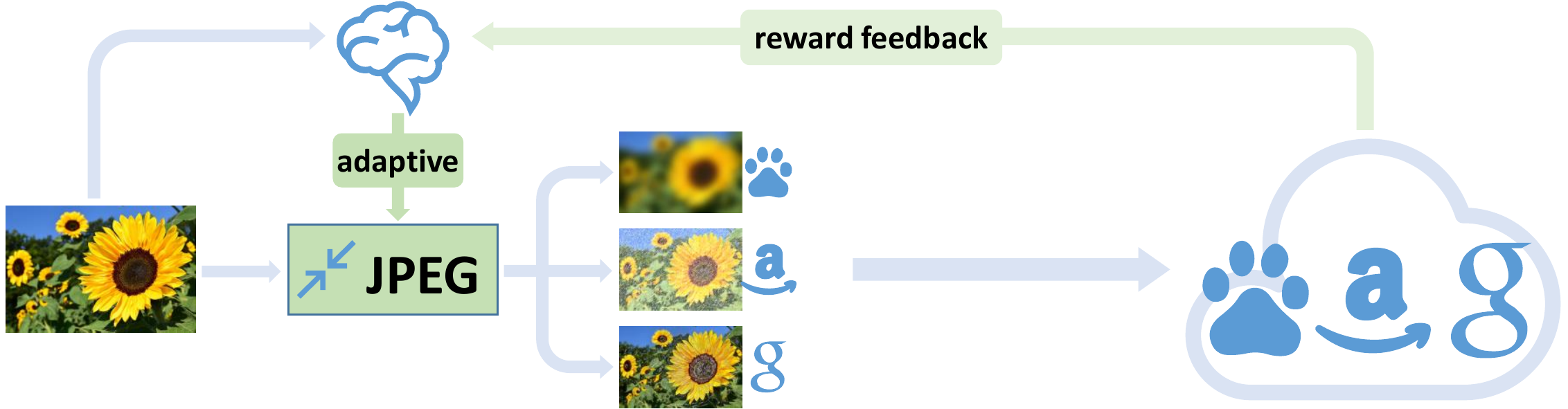}}
		\begin{center}
			{(b) Adaptive solution: input image and backend model aware compression}
		\end{center}
	\end{minipage}
	\caption{Comparing to the conventional solution, my solution updates the compression strategy based on the feedback from the backend model}
	\label{fig: framework}
\end{figure}

A conceptual framework of the solution is shown in Fig.~\ref{fig: framework}. Briefly, it is a deep reinforcement learning-based system to train an agent to choose the proper quality level $c$ for an image to be compressed in JPEG format. Note that the Deep Q-learning network-based agent's behaviors are various for different input image ``sceneries'' and backend cloud services. By analyzing the agent's behaviors using, they provided the reasons that the agent chooses a specific compression quality level, and revealed that images containing large smooth areas are more sensitive to compression, while images with complex textures are more robust to compression for computer vision models. The full design of AdaCompress\footnote{AdaCompress works with online computer vision-based APIs \url{https://github.com/hosea1008/AdaCompress}} is then able to choose a proper compression quality level for an image to a computer vision model on the cloud-end, in an online manner.

\section{Learning-based Edge Content Serving} \label{sec:caching}

\subsection{Mobility-driven Edge Content Caching}

Video clips are increasingly being generated by users and instantly shared with their friends. In contrast to conventional live and on-demand video streaming that are consumed using TVs and PCs, mobile video streaming is generally watched by users on mobile devices with wireless connections, \emph{i.e.}, 4G/5G cellular or Wi-Fi. User behaviors and wireless network quality in mobile video streaming \cite{hu2014quality,liang2015wireless} can be quite different from those in conventional video streaming \cite{adhikari2012unreeling,mukerjee2015enabling}, thus requiring improvements in the delivery of mobile video streaming.

To meet the sky-rocketing increase in bandwidth requirements resulting from data-intensive video streaming and to reduce the monetary cost for renting expensive resources in conventional content delivery networks (CDNs), video service providers are pushing their content delivery infrastructure closer to users to utilize network and storage resources in households for content delivery \cite{baochun-tomccap-streaming2013}, including caching content over femtocells \cite{golrezaei2012femtocaching} and replicating video content via Wi-Fi smartrouters in households. Youku, one of the largest online video providers in China, has deployed over $300$K smartrouters in its users’ homes in less than one year, expecting to transform a large fraction of its users ($250$M) into such content delivery peer nodes \cite{Ming-nossdav16}. To serve users with good quality of experience using the new edge network solutions, it is important to answer the following questions: 1) What are the video request patterns in mobile video streaming? 2) How do users behave in today's mobile video systems, and what is the implication of their behaviors on edge network video content delivery? 3) How is the quality of user experience in mobile video sessions? 4) Can today's mobile network infrastructure appropriately satisfy the mobile video streaming demand? 5) What strategies can be applied to best support mobile video content delivery?

Several measurement studies have been conducted to address the above questions. However, such measurement studies are challenging because many different factors are involved, including user behaviors (\emph{i.e.}, mobility pattern and video preference), video content characteristics, and mobile network characteristics. Previous studies generally focus on a single aspect, \emph{e.g.}, studying the popularity of mobile video content \cite{brodersen2012youtube,xu2015forecasting}, user mobility behaviors \cite{das2014contextual}, or network strategies to support mobile video streaming, \emph{e.g.}, content replication \cite{gitzenis2013asymptotic}. The limitation of the previous studies is that they have not considered the joint impact of user behaviors, content characteristics, and wireless network deployment, on edge network content delivery.

The above questions are addressed from the perspectives of both the mobile video service and wireless network providers. From the perspective of mobile video service, how users view mobile videos, including their mobility patterns in video sessions and the content selection in different locations, are studied to build a mobile video consumption model. From the perspective of wireless network provider, the mobile video requests can be served by both the Wi-Fi and cellular infrastructures that are commonly used by today's users, and there are analytical insights on how to improve the QoS of wireless networks according to their video request patterns.

Based on previous results, both mobility and geographic migration behaviors of users can significantly affect mobile video requests. In particular, the mobility behaviors of users are heterogeneous, \emph{e.g.}, a number of multi-location users request videos intensively and request them in different locations, whereas there is a large fraction of users who only request a small number of videos in the same location. For the geographic migration behaviors, users have regular commute behaviors, involving $2$--$3$ regularly visited locations where they tend to request mobile videos, and it is common for users to move between the same type of locations (\emph{e.g.}, residential) and issue video requests. \emph{These observations suggest that joint caching strategies over multiple locations can improve the caching performance}.

It is then reasonable to compare the effectiveness of Wi-Fi and cellular-based edge network caching solutions, and study the potential improvement on mobile video streaming to today's wireless networks. Based on the edge network traces covering $1,055,881$ Wi-Fi APs and $69,210$ cellular base stations, Ma \emph{et al.}~\cite{ge-jsac2017} investigated conventional caching strategies, including least recently used (LRU) and least frequently used (LFU) for edge network mobile video delivery. Most of today's Wi-Fi and cellular deployments are close enough to the mobile requests of users in different locations; however, although Wi-Fi and cellular have different deployment strategies, they cannot well serve different categories of mobile video users. Second, a number of factors including user mobility, content popularity, cache capacity, and caching strategies affect the caching performance for both Wi-Fi and cellular caching for mobile video delivery. For example, unpopular videos attract users mostly from few locations where users have particular interests in the content, and caching strategies have various influences on different categories of users.

Motivated by the measurement insights, a geo-collaborative caching strategy can be designed for mobile video delivery, which jointly considers mobile video request patterns, user behaviors and the deployment of wireless networks. According to real-world trace-driven experiments, such design is able to achieve a $20\%$ (\emph{resp.} $30\%$) cache hit rate improvement and a $20\%$ (\emph{resp.} $30\%$) service rate improvement compared with conventional LRU (\emph{resp.} LFU) caching strategies.

\subsection{Joint Mobility and Social Content Replication}

Mobile social network services based on the convergence of wireless networks, smart devices, and online social networks have witnessed rapid expansion in recent years \cite{zhangunderstand}. According to YouTube, over $100$ hours worth of videos have been produced by individuals and shared among themselves, and the traffic resulting from delivering these content items to mobile devices has exceeded $50\%$, significantly challenging the traditional content delivery paradigm, in which content is replicated by a hierarchical infrastructure using the same scheme. It is usually expansive and inefficient to replicate the massive number of social content items to traditional CDN servers.

With the development of \emph{device-to-device} communication, it has become promising to offload the bandwidth-intensive social content delivery to users' mobile devices and let them serve each other. Previous studies have demonstrated that such device-to-device content sharing is possible when users are close to each other and when the content to be delivered is delay tolerant.  \emph{Mobile edge networks} (or edge networks for short)  define the local area whereby users move across regions and can directly communicate with each other. It is intriguing to investigate content delivery strategies in the context of edge networks because both users' behaviors and network properties have to be studied.

In traditional device-to-device content sharing, a user typically sends the generated content to a set of users that are close to the user in a broadcast-like manner, therein causing the following problems: 1) Due to the broadcasting mechanism, users' devices have to expend high amounts of power to cache and forward many content items in the edge network. As the number of user-generated social content items increases, such a mechanism becomes inherently in-scalable. 2) Social content---due to the dynamical social propagation---has heterogeneous popularity, whereas conventional approaches treat all such content the same, resulting in wasted resources to replicate unpopular content items. 3) Due to the dynamic mobility patterns, it is difficult to guarantee any quality of experience.

To address these problems, a joint propagation- and mobility-aware replication strategy \cite{wang2016propagation} is proposed based on social propagation characteristics and user mobility patterns in the edge-network \emph{regions}, {\em e.g.}, $100\times100$m$^2$ areas that users can move across and deliver content to other users. The idea behind the proposal is as follows. 1) Instead of letting content flood between users that are merely close to one another, it is reasonale to replicate social content according to the social influence of users and the social propagation of content. 2) A regional social popularity prediction model is proposed to capture the popularity of content items based on both regional and social information. 3) Then, the system can replicate social content items according to not only regional social popularity but also user mobility patterns, which capture how users move across and remain in these edge-network regions.

The whole framework is facing the following challenges: How does one capture the joint propagation and mobility behaviors? How does one identify the parameters that affect the performance of mobile social content replication? How does one design efficient strategies/algorithms for the proposal that work in the real world? 

Social propagation and user mobility predictive models are designed to capture the popularity distribution of content in different regions. Using the predictive models, the D2D content replication can be formulated as an optimization problem, which is inherently centralized. It can be solved by a heuristic algorithm in a distributed manner practically, therein only requiring historical, local and partial information.

In an online social network, users share content with their friends through social connections. Both social graphs and user behaviors determine the propagation of contents. Because such social graphs and user behaviors are inside the online social network, they can be independent of users' mobility patterns in the physical world. For example, a user can intensively interact with their friends online without having to be at the same location thanks to the online social network.

\begin{figure}[t]
	\centering
		\includegraphics[width=.8\linewidth]{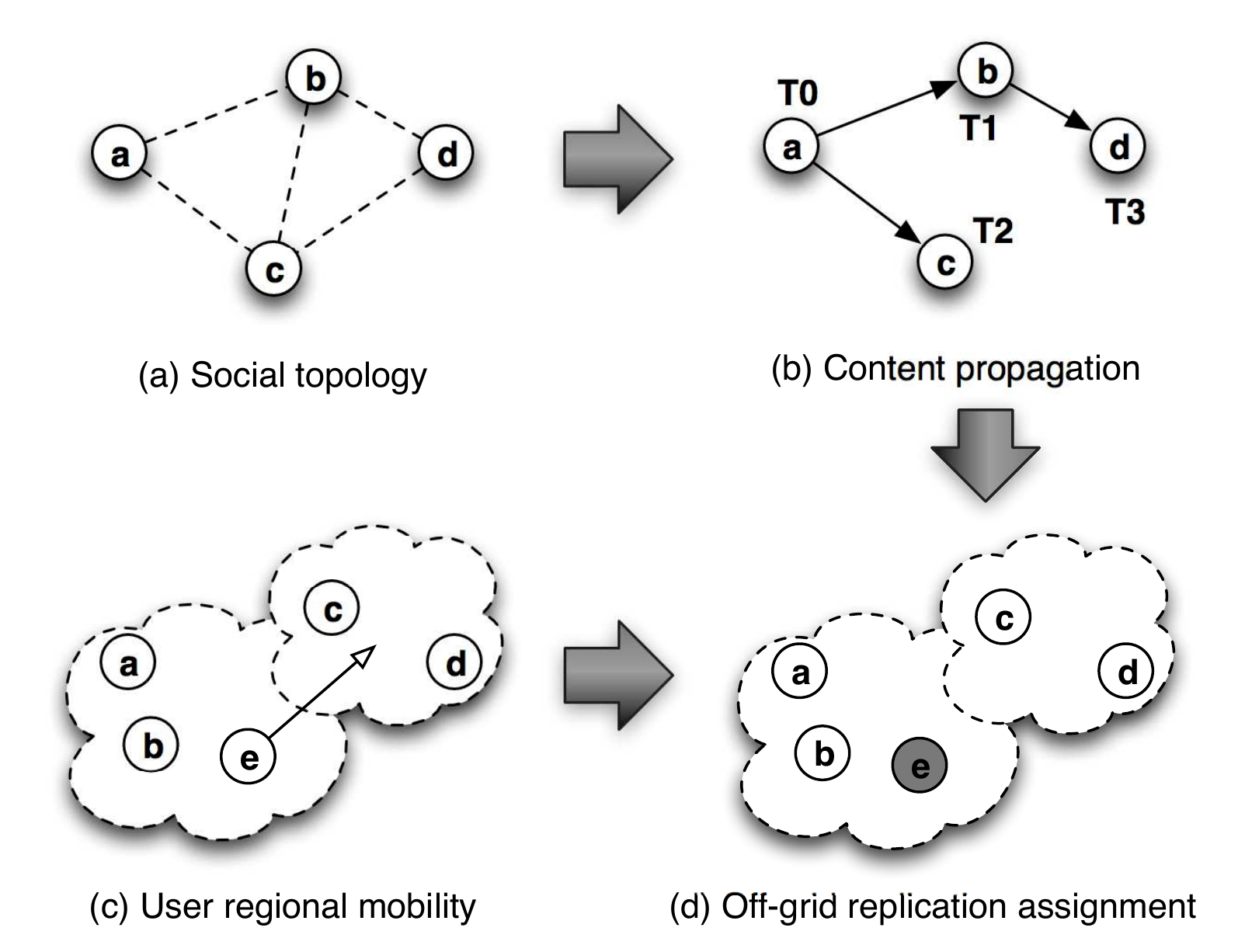}
		\caption{D2D replication affected by social topology, content propagation and user regional mobility.}
	\label{fig:user-topologies}
\end{figure}

As illustrated in Fig.~\ref{fig:user-topologies}, based on the social graph and propagation patterns, how contents will be received by users can be estimate; based on the regional mobility, which regions users will be moving to and how long they will stay are predictable. Simultaneously, which users will replicate which social contents on the move can be decided. In this example, user $e$ -- while not a friend of any other user -- is moving to the region where user $c$ and $d$ are located. Thus, $e$ will be selected to replicate the content generated by user $a$, and both user $c$ and user $d$ will receive the content shared by user $a$ in the social propagation at times $T2$ and $T3$, respectively.

So, it is reasonable to jointly utilize the social graph, user behaviors and user mobility patterns for D2D replication to serve edge-network regions. To this end, the propagation of mobile social contents, user mobility patterns, and characteristics of edge-network regions are considered in one framework. Based on the measurement insights, a propagation- and mobility-aware D2D replication for edge-network regions is proposed. It decouples the content replication for the social propagation on the online social network from users' mobility in the physical world, \emph{i.e.}, a user may cache content and deliver the content to other users who are not socially connected to the user. Note that although the replication is decoupled from the mobility, the social propagation and mobility patterns do not have to be independent, \emph{e.g.}, users may share more content at some locations compared to other locations. The design provides an algorithm to select content for users to cache adaptively. The solution improves the D2D delivery fraction by $4$ times compared to a pure movement-based approach and by $2$ times compared to a pure popularity-based approach.

\subsection{Edge Caching with Joint Cache Size Scaling and Replacement Adaptation}

Many vertical video-sharing platforms have to operate their own video streaming due to their unique interactive requirements (\emph{e.g.}, online education platform need to incorporate quiz in video streaming) that cannot be well handled by conventional large video sharing platforms (\emph{e.g.}, YouTube). Such small content providers can attract a large number of users driving significant internet traffic. Compared with conventional large content providers that maintain their own content distribution infrastructure globally, small content providers tend to lease resources from the emerging elastic content delivery networks, \emph{e.g.}, Akamai Aura \cite{akamai_aura} and Huawei uCDN \cite{huawei_ucdn}, for their content delivery. Such ECDNs provide their tenants ``content delivery as a service'' with \emph{elastic} configurations for the content distribution, not only the primary content assignment to peering servers but also \emph{scaling cache size} and \emph{changing the content replacement strategies} \emph{jointly} and \emph{dynamically}, for more flexibility along with their business expansion.

Using cache has been a design philosophy to improve performance for several systems over the decades. However, regarding the strategies for a cache system, scaling the cache infrastructure and changing the replacement strategy are only separately studied: adjusting cache size while fixing the replacement strategy, \emph{e.g.}, \cite{kwak2018dynamic, Dehghan_sharingLRU, Chu_onallocating}; or conversely, shifting content replacement strategy while fixing the cache size, \emph{e.g.}, \cite{lee2001lrfu, lee1999existence, megiddo2003arc}.

Previous studies usually have the following limitations when applied in elastic content delivery networks scenario. First, being able to jointly tune the cache size (representing the infrastructure scale) and replacement strategies can a content provider achieve the maximum utility in content delivery, and this makes the studies that can only tune either cache replacement strategy or cache size separately, less efficient if not totally ineffective for small content providers to utilize elastic content delivery networks. Second, previous studies usually ignore the cost of renting cache servers, which is essential, especially for small content providers with a much tighter budget for content distribution; both the cost and cache performance determine the final utility experienced by a small content provider.

To this end, Ye \emph{et al.}~\cite{ye2021infocom} proposed to design an elastic caching framework based on deep reinforcement learning that jointly and dynamically adjusts cache size scaling and content replacement strategy. The rationale is that deep reinforcement learning is inherently suited to solving sequential decision problems since it optimizes long-term reward \cite{mnih2015human, silver2014deterministic}. However, designing a joint elastic caching framework will face two significant challenges, which cannot be addressed by existing dynamic caching methods: \begin{itemize}

	\item \emph{Fast and dynamical content popularity changes.} Small content providers usually have non-stationary and fast-changing content request patterns; Meanwhile, elastic content delivery network providers typically charge the small content providers using dynamical pricing schemes (\emph{e.g.}, spot pricing), which further makes the utility of content providers dynamically change over time. These have challenged conventional strategy assuming a stationary popularity pattern and stable pricing.
	
	\item \emph{Actions to be jointly determined are usually from different decision spaces.} A joint elastic caching framework requires to decide two action variables simultaneously. Cache size scaling ($a_1$) is a \emph{discrete} (\emph{e.g.}, an integer) or \emph{continuous} variable (\emph{e.g.}, a real number); while, content replacement strategy selection ($a_2$) is usually a \emph{categorical} variable (\emph{i.e.}, no distance measure between two variable). Nevertheless, existing deep reinforcement learning algorithms suffer performance degradation when there are discrete variables, which is called ``discrete challenge'' in this paper.

\end{itemize}

\begin{figure}[!t]
     \begin{minipage}[t]{.90\linewidth}
          \centering
               \includegraphics[width=\linewidth]{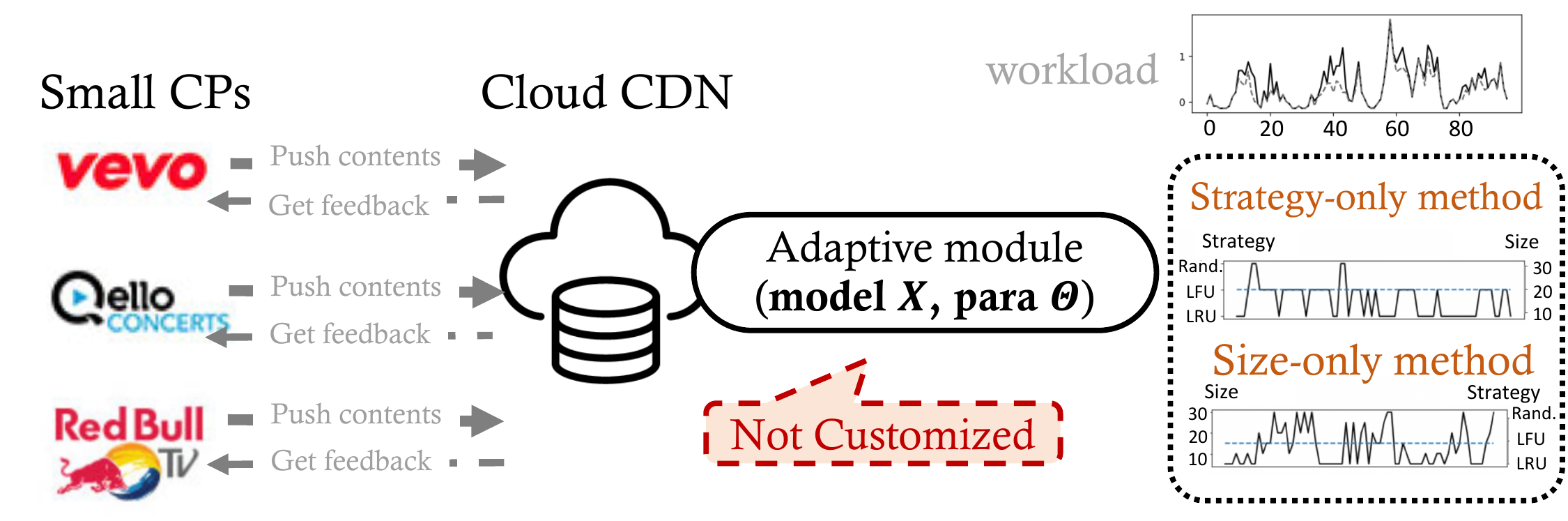}
          \centerline{\scriptsize (a) Traditional adaptive caching solutions in cloud CDN.}
     \end{minipage}
     \hfill
     \begin{minipage}[t]{.90\linewidth}
          \centering
               \includegraphics[width=\linewidth]{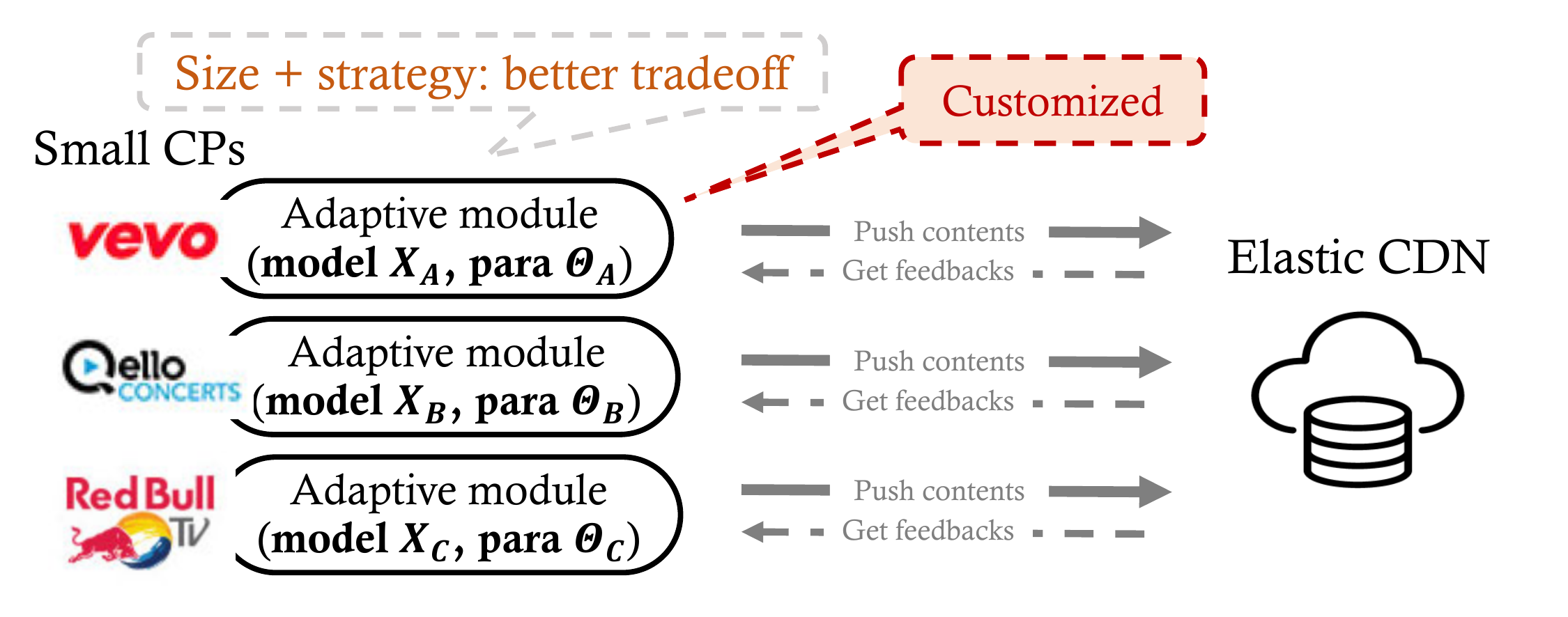}
          \centerline{\scriptsize (b) The joint adaptive caching solution for small content providers.}
     \end{minipage}
     \caption{Comparison between previous studies on adaptive caching and the joint adaptive caching solution.}
	\label{fig:motivation}
\end{figure}

To address these challenges and provide a cost-effective method for small content providers to efficiently use elastic content delivery networks, a distribution-guided deep reinforcement learning-based framework called JEANA (\underline{J}oint \underline{E}lastic c\underline{A}chi\underline{N}g decision m\underline{A}ker) is proposed. The ultimate goal is to maximize a small content provider's long-term utility, which comprises of the performance gain (e.g., cache hit rate) of satisfying requests at the cache, minus the rental cost of cache service. Fig.~\ref{fig:motivation} shows the uniqueness of the proposal.

First, to address the challenge caused by the fast and dynamical content popularity changes, JEANA uses a model-free DRL framework to learn a policy network to generate a joint caching strategy without prior information of content popularity and dynamic pricing. In particular, a \emph{policy network} is designed to combine long-term, short-term request patterns and current cache status as input to detect and adapt to rapid request pattern changes. An \emph{advantage actor-critic} model is employed to keep the training process stable, while constraining the policy update step by a KL-divergence to provide the policy \emph{improvement guarantee}.

Second, to handle the joint actions that are from different decision spaces, JEANA maintains \emph{structural} information among action variables when handling discrete control problems. Specifically, the cache performance curve versus either cache size or replacement strategy is \emph{unimodal}; based on this insight, a \emph{discrete normal distribution} based policy and a Wasserstein distance loss are designed to constrain the action probability distribution to be smoother. It is proved that the distribution-guided deep reinforcement learning method has policy \emph{improvement guarantee}.

\subsection{Incentive and Fairness in Edge Content Caching}

Enabled by the proliferation of mobile devices and the advent of dedicated mobile video applications, mobile video traffic has already become the largest mobile traffic category. To improve the content access quality experienced by users and alleviate the load for both CDN servers and backbone networks, CDN service providers are moving content distribution capacity to the edge networks (\emph{e.g.}, on access points). Edge-network based content delivery platform acts as a crowdsourcing system that can offload content distribution tasks to massive edge devices \cite{golrezaei2012femtocaching, chen2015thunder, ma2016understanding}, forming what we called \emph{crowdsourced CDN}.

This solution becomes promising thanks to the emergence of smart access points (\emph{e.g.}, femtocell \cite{golrezaei2012femtocaching}, smartrouters \cite{ma2016understanding}) equipped with an OS and storage devices (\emph{e.g.}, a hard disk or an SD card). For example, Akamai's Intelligent Platform leverages millions of users' devices to assist its content delivery infrastructure. Moreover, recent studies show that utilizing the spare resources of edge access points (APs) in crowdsourced CDN can achieve better overall system performance, \emph{e.g.}, reducing the cost of content providers and the content access latency of users \cite{gharaibeh2016provably, ma2017aprank}. Nevertheless, the move to crowdsourced CDN is still facing the following challenges.

First, multiple content providers (CPs) competing for the scarce crowdsourced CDN resources result in unfairness and low overall utilization \cite{douros2017caching}. In the crowdsourced CDN scenario where individuals' contributions are scarce resources, CPs tend to use false (usually higher) bidding prices to compete for crowdsourced resources. Guaranteeing fair resource allocation and truthful bidding is challenging, especially when we consider the unique setup in crowdsourced CDN as follows. 1) It is common for CPs to compete for resources in the same edge region or the same time period, resulting in high resource competition, particularly when the resources are limited. 2) The resource demand of each CP is dynamically changing over time and across different geographical regions, making static resource bidding strategies \cite{Wang2017Towards} ineffective.

Second, though \emph{edge-network owners} (\emph{i.e.}, owners of edge APs) contributing resources can be rewarded by crowdsourced CDN, the supply of resources is highly volatile. 1) The resource supply is dynamically changing over time, because owners are free to decide whether and how much they are willing to contribute the resource individually, making conventional static pricing strategies \cite{mohsenianrad2010autonomous} ineffective. 2) The distribution of resource supply is highly skewed across different regions \cite{ma2017understanding}. 3) The resource supply is highly sensitive to the change in prices (intuitively, higher price leads to more resource supply but also results in higher cost) \cite{song2014energy}. It is thus challenging to design incentive strategies that maintain a sustainable resource supply for crowdsourced CDN.

To address the challenges above, based on the game and economics theory, we propose a joint incentive and fairness guaranteed solution, which not only optimally matches the owners looking to contribute their resources with CPs looking to rent them, but also fairly decides the allocation of limited resources among multiple CPs. Conventional strategies can only follow \emph{fixed} rules to collect resources from owners and rent them out to CPs, while the design focuses on dynamic resource collection and allocation in a \emph{random} environment, as well as formulating games based on public information rather than specific private information. The design targets the following properties. a) Fairness: the limited resources can be fairly allocated to CPs. b) Utility maximization: the design maximizes the aggregate social welfare (\emph{i.e.}, aggregate utility of CPs, owners and the crowdsourced CDN service provider). c) Truthfulness: declared true demand for resources is a dominant strategy for each CP. d) Computational efficiency: the resource allocation and pricing algorithms execute in polynomial time. 

Large-scale measurement studies have been performed on real-world operational traces to understand the spatial and temporal characteristics of content requests in bandwidth-intensive mobile applications. The observations reveal that the resources contributed by individuals can be highly limited and volatile, if not well incentivized. Meanwhile, there exists high resource competition among CPs, with heterogeneous demands in different geographical regions. The measurement studies also provide principles for the fairness and incentive mechanism design, \emph{e.g.}, diverse roles that CPs play (\emph{i.e.}, their traffic characteristics and different resource demands) are taken into account.

A joint incentive and fairness guaranteed solution is proposed for resource \emph{collection} (from owners) and \emph{allocation} (to CPs) in a crowdsourced CDN platform \cite{ma2018wireless}. In the crowd resource collection, there is a problem of matching between the volatile crowdsourced supply and ever-changing demand from CPs. To solve this problem in a distributed and automatic manner, a Stackelberg game is designed to enable ``bargain'' between resource owners and the crowdsourced CDN service provider, and prove the existence of a Stackelberg equilibrium, achieving the optimal incentivized price. By coding the rewarding price, a genetic algorithm to reach the equilibrium in polynomial time is proposed.

In the fairness-guaranteed resource allocation, CPs are regarded as players in a coalitional game and prove that CPs are willing to form a grand coalition to compete for resources. a Shapley-based algorithm resolves resource allocation across different regions, which satisfies efficiency, symmetry, dummy, and additivity properties. To guarantee the rationality of the fair resource allocation, a heterogeneous charging scheme is used for CPs with heterogeneous demands. The design is able to not only handle CPs' competition but also enable truthfulness of CPs with different demands to only bid according to their true demands.

\section{Concluding Remarks} \label{sec:conclusion}

In this article, we review the emerging research topic of edge multimedia computing. We present the latest results on edge sensing, first-mile content harvesting and strategical edge re-routing; we show scalable machine learning inference using both model decoupling and model compression, with the assistance of edge infrastructure; and we present mobility and social driven edge content replication. These studies show that edge computing not only provides the essential infrastructure for improving multimedia services but also drive innovations for new edge-aware multimedia.

Almost a decade ago multimedia cloud computing was introduced to address the migration from the multimedia-aware cloud (media cloud) and cloud-aware multimedia (cloud media) \cite{zhu2011multimedia}. It shows how a cloud could perform distributed multimedia processing and storage and provide quality of service provisioning for multimedia services. Today, the new trend is that multimedia data is collected, processed, stored, and served more and more at the edge network using the edge computing infrastructure. How to migrate conventional multimedia workflows to the new edge context, is of great importance to both industry and academia. We have seen exciting development towards edge-aware multimedia and multimedia-aware edge compupting that jointly would enable many new applications in multimedia edge computing, with the hope of making high-quality, scalable, personalized, and privacy-protected multimedia services to edge-networked users a reality.

\bibliographystyle{plain}
\bibliography{edge.bib}

\begin{thebibliography}{10}

\bibitem{akamai_aura}
Akamai collaborates with orange on nfv initiative to dynamically scale cdn
  capacity for large event.
\newblock In {\em [Online]. Available:
  https://www.akamai.com/us/en/about/news/press/2016-press/akamai-collaborates-with-orange-on-nfv-initiative.jsp}.

\bibitem{abbas2017mobile}
Nasir Abbas, Yan Zhang, Amir Taherkordi, and Tor Skeie.
\newblock Mobile edge computing: A survey.
\newblock {\em IEEE Internet of Things Journal}, 5(1):450--465, 2017.

\bibitem{adhikari2012unreeling}
V.~K. Adhikari, Y.~Guo, F.~Hao, M.~Varvello, V.~Hilt, M.~Steiner, and Z.~Zhang.
\newblock Unreeling netflix: Understanding and improving multi-cdn movie
  delivery.
\newblock In {\em INFOCOM, 2012 Proceedings IEEE}, pages 1620--1628. IEEE,
  2012.

\bibitem{brodersen2012youtube}
A.~Brodersen, S.~Scellato, and M.~Wattenhofer.
\newblock Youtube around the world: geographic popularity of videos.
\newblock In {\em Proceedings of the 21st international conference on World
  Wide Web}, pages 241--250. ACM, 2012.

\bibitem{chen2015thunder}
L.~Chen, Y.~Zhou, M.~Jing, and R.~TB Ma.
\newblock Thunder crystal: a novel crowdsourcing-based content distribution
  platform.
\newblock In {\em Proceedings of the 25th ACM Workshop on Network and Operating
  Systems Support for Digital Audio and Video (NOSSDAV)}, pages 43--48. ACM,
  2015.

\bibitem{chen2015glimpse}
Tiffany Yu-Han Chen, Lenin Ravindranath, Shuo Deng, Paramvir Bahl, and Hari
  Balakrishnan.
\newblock Glimpse: continuous, real-time object recognition on mobile devices.
\newblock In {\em 13th ACM Conference on Embedded Networked Sensor Systems},
  pages 155--168. ACM, 2015.

\bibitem{Chu_onallocating}
Weibo Chu, Mostafa Dehghan, Don Towsley, and Zhi-li Zhang.
\newblock {On Allocating Cache Resources to Content Providers}.
\newblock In {\em Proceedings of the 3rd ACM Conference on Information-Centric
  Networking}, pages 154--159, 2016.

\bibitem{das2014contextual}
A.~K. Das, P.~H. Pathak, C.~Chuah, and P.~Mohapatra.
\newblock Contextual localization through network traffic analysis.
\newblock In {\em INFOCOM, 2014 Proceedings IEEE}, pages 925--933. IEEE, 2014.

\bibitem{Dehghan_sharingLRU}
Mostafa Dehghan, Weibo Chu, Philippe Nain, and Don Towsley.
\newblock {Sharing LRU Cache Resources among Content Providers : A
  Utility-Based Approach}.
\newblock {\em IEEE/ACM Transactions on Networking}, 27(2):477--490, 2019.

\bibitem{delac2005effects}
Kresimir Delac, Mislav Grgic, and Sonja Grgic.
\newblock Effects of jpeg and jpeg2000 compression on face recognition.
\newblock In {\em International Conference on Pattern Recognition and Image
  Analysis}, pages 136--145. Springer, 2005.

\bibitem{douros2017caching}
V.~G Douros, S.~E. Elayoubi, E.~Altman, and Y.~Hayel.
\newblock Caching games between content providers and internet service
  providers.
\newblock {\em Performance Evaluation}, 113:13--25, 2017.

\bibitem{evtimov2018robust}
Ivan Evtimov, Kevin Eykholt, Earlence Fernandes, Tadayoshi Kohno, Bo~Li, Atul
  Prakash, Amir Rahmati, and Dawn Song.
\newblock Robust physical-world attacks on deep learning models.
\newblock In {\em Computer Vision and Pattern Recognition}, 2018.

\bibitem{finn2017model}
Chelsea Finn, Pieter Abbeel, and Sergey Levine.
\newblock Model-agnostic meta-learning for fast adaptation of deep networks.
\newblock In {\em Proceedings of the 34th International Conference on Machine
  Learning-Volume 70}, pages 1126--1135. JMLR. org, 2017.

\bibitem{garg2019online}
Navneet Garg, Mathini Sellathurai, Vimal Bhatia, BN~Bharath, and Tharmalingam
  Ratnarajah.
\newblock Online content popularity prediction and learning in wireless edge
  caching.
\newblock {\em IEEE Transactions on Communications}, 68(2):1087--1100, 2019.

\bibitem{gharaibeh2016provably}
A.~Gharaibeh, A.~Khreishah, B.~Ji, and M.~Ayyash.
\newblock A provably efficient online collaborative caching algorithm for
  multicell-coordinated systems.
\newblock {\em IEEE Transactions on Mobile Computing}, 15(8):1863--1876, 2016.

\bibitem{gitzenis2013asymptotic}
S.~Gitzenis, G.~Paschos, and L.~Tassiulas.
\newblock Asymptotic laws for joint content replication and delivery in
  wireless networks.
\newblock {\em Information Theory, IEEE Transactions on}, 59(5):2760--2776,
  2013.

\bibitem{golrezaei2012femtocaching}
N.~Golrezaei, K.~Shanmugam, A.~G Dimakis, A.~F Molisch, and G.~Caire.
\newblock Femtocaching: Wireless video content delivery through distributed
  caching helpers.
\newblock In {\em Proceedings of the 31st IEEE International Conference on
  Computer Communications (INFOCOM)}, pages 1107--1115. IEEE, 2012.

\bibitem{gueguen2018faster}
Lionel Gueguen, Alex Sergeev, Ben Kadlec, Rosanne Liu, and Jason Yosinski.
\newblock Faster neural networks straight from jpeg.
\newblock In {\em Advances in Neural Information Processing Systems}, pages
  3933--3944, 2018.

\bibitem{Res_Net}
Kaiming He, Xiangyu Zhang, Shaoqing Ren, and Jian Sun.
\newblock Deep residual learning for image recognition.
\newblock In {\em IEEE CVPR}, pages 770--778, 2016.

\bibitem{hu2014quality}
W.~Hu and G.~Cao.
\newblock Quality-aware traffic offloading in wireless networks.
\newblock In {\em Proceedings of the 15th ACM international symposium on Mobile
  ad hoc networking and computing}, pages 277--286. ACM, 2014.

\bibitem{huawei_ucdn}
Huawei.
\newblock Huawei ucdn solution.
\newblock In {\em [Online]. Available: http://carrier.
  huawei.com/en/solutions/cloud-powered-digital-services/ucdn}.

\bibitem{jain2015overlay}
Puneet Jain, Justin Manweiler, and Romit Roy~Choudhury.
\newblock Overlay: practical mobile augmented reality.
\newblock pages 331--344. ACM, 2015.

\bibitem{jiang2018chameleon}
Junchen Jiang, Ganesh Ananthanarayanan, Peter Bodik, Siddhartha Sen, and Ion
  Stoica.
\newblock Chameleon: scalable adaptation of video analytics.
\newblock In {\em Proceedings of the 2018 Conference of the ACM Special
  Interest Group on Data Communication}, pages 253--266, 2018.

\bibitem{kang2017neurosurgeon}
Yiping Kang, Johann Hauswald, Cao Gao, Austin Rovinski, Trevor Mudge, Jason
  Mars, and Lingjia Tang.
\newblock Neurosurgeon: Collaborative intelligence between the cloud and mobile
  edge.
\newblock {\em ACM SIGARCH Computer Architecture News}, 45(1):615--629, 2017.

\bibitem{konevcny2016federated}
Jakub Kone{\v{c}}n{\`y}, H~Brendan McMahan, Felix~X Yu, Peter Richt{\'a}rik,
  Ananda~Theertha Suresh, and Dave Bacon.
\newblock Federated learning: Strategies for improving communication
  efficiency.
\newblock {\em arXiv preprint arXiv:1610.05492}, 2016.

\bibitem{kwak2018dynamic}
Jeongho Kwak, Georgios Paschos, and George Iosifidis.
\newblock Dynamic cache rental and content caching in elastic wireless cdns.
\newblock In {\em 2018 16th International Symposium on Modeling and
  Optimization in Mobile, Ad Hoc, and Wireless Networks (WiOpt)}, pages 1--8.
  IEEE, 2018.

\bibitem{lee1999existence}
Donghee Lee, Jongmoo Choi, Jong-Hun Kim, Sam~H Noh, Sang~Lyul Min, Yookun Cho,
  and Chong-Sang Kim.
\newblock On the existence of a spectrum of policies that subsumes the least
  recently used (lru) and least frequently used (lfu) policies.
\newblock In {\em SIGMETRICS}, volume~99, pages 1--4. Citeseer, 1999.

\bibitem{lee2001lrfu}
Donghee Lee, Jongmoo Choi, Jong-Hun Kim, Sam~H Noh, Sang~Lyul Min, Yookun Cho,
  and Chong~Sang Kim.
\newblock Lrfu: A spectrum of policies that subsumes the least recently used
  and least frequently used policies.
\newblock {\em IEEE transactions on Computers}, (12):1352--1361, 2001.

\bibitem{baochun-tomccap-streaming2013}
B.~Li, Z.~Wang, J.~Liu, and W.~Zhu.
\newblock Two decades of internet video streaming: A retrospective view.
\newblock {\em ACM transactions on multimedia computing, communications, and
  applications (TOMM)}, 9(1s):33, 2013.

\bibitem{li2019adacompress}
Hongshan Li, Yu~Guo, Zhi Wang, Shutao Xia, and Wenwu Zhu.
\newblock Adacompress: Adaptive compression for online computer vision
  services.
\newblock In {\em Proceedings of the 27th ACM International Conference on
  Multimedia}, pages 2440--2448, 2019.

\bibitem{li2018jalad}
Hongshan Li, Chenghao Hu, Jingyan Jiang, Zhi Wang, Yonggang Wen, and Wenwu Zhu.
\newblock Jalad: Joint accuracy-and latency-aware deep structure decoupling for
  edge-cloud execution.
\newblock In {\em 2018 IEEE 24th International Conference on Parallel and
  Distributed Systems (ICPADS)}, pages 671--678. IEEE, 2018.

\bibitem{liang2015wireless}
C.~Liang and F.~R. Yu.
\newblock Wireless network virtualization: A survey, some research issues and
  challenges.
\newblock {\em Communications Surveys \& Tutorials, IEEE}, 17(1):358--380,
  2015.

\bibitem{DeepN-JPEG}
Zihao Liu, Tao Liu, Wujie Wen, Lei Jiang, Jie Xu, Yanzhi Wang, and Gang Quan.
\newblock Deepn-jpeg: a deep neural network favorable jpeg-based image
  compression framework.
\newblock In {\em Proceedings of the 55th Annual Design Automation Conference},
  page~18. ACM, 2018.

\bibitem{ma2017aprank}
G.~Ma, Z.~Wang, M.~Chen, and W.~Zhu.
\newblock Aprank: Joint mobility and preference-based mobile video prefetching.
\newblock In {\em Multimedia and Expo (ICME), 2017 IEEE International
  Conference on}, pages 7--12. IEEE, 2017.

\bibitem{ma2018wireless}
Ge~Ma, Zhi Wang, Jiahui Ye, and Wenwu Zhu.
\newblock Wireless caching in large-scale edge access points: A local
  distributed approach.
\newblock In {\em Proceedings of the 24th Annual International Conference on
  Mobile Computing and Networking}, pages 726--728, 2018.

\bibitem{ma2017understanding}
Ge~Ma, Zhi Wang, Miao Zhang, Jiahui Ye, Minghua Chen, and Wenwu Zhu.
\newblock Understanding performance of edge content caching for mobile video
  streaming.
\newblock {\em IEEE Journal on Selected Areas in Communications},
  35(5):1076--1089, 2017.

\bibitem{ge-jsac2017}
Ge~Ma, Zhi Wang, Miao Zhang, Jiahui Ye, Minghua Chen, and Wenwu Zhu.
\newblock {Understanding Performance of Edge Content Caching for Mobile Video
  Streaming}.
\newblock In {\em IEEE Journal on Selected Areas in Communications (JSAC)},
  2017.

\bibitem{Ming-nossdav16}
M.~Ma, Z.~Wang, K.~Su, and L.~Sun.
\newblock {Understanding Content Placement Strategies in Smartrouter-based Peer
  Video CDN}.
\newblock In {\em ACM SIGMM Workshop on Network and Operating Systems Support
  for Digital Audio and Video (NOSSDAV)}, 2016.

\bibitem{ma2016understanding}
M.~Ma, Z.~Wang, K.~Su, and L.~Sun.
\newblock Understanding content placement strategies in smartrouter-based peer
  video cdn.
\newblock In {\em Proceedings of the 26th ACM International Workshop on Network
  and Operating Systems Support for Digital Audio and Video (NOSSDAV)}, page~7.
  ACM, 2016.

\bibitem{mach2017mobile}
Pavel Mach and Zdenek Becvar.
\newblock Mobile edge computing: A survey on architecture and computation
  offloading.
\newblock {\em IEEE Communications Surveys \& Tutorials}, 19(3):1628--1656,
  2017.

\bibitem{mao2017mobile}
Yuyi Mao, Changsheng You, Jun Zhang, Kaibin Huang, and Khaled~B Letaief.
\newblock Mobile edge computing: Survey and research outlook.
\newblock {\em arXiv preprint arXiv:1701.01090}, 2017.

\bibitem{megiddo2003arc}
Nimrod Megiddo and Dharmendra~S Modha.
\newblock Arc: A self-tuning, low overhead replacement cache.
\newblock In {\em FAST}, volume~3, pages 115--130, 2003.

\bibitem{DQN}
Volodymyr Mnih, Koray Kavukcuoglu, David Silver, Alex Graves, Ioannis
  Antonoglou, Daan Wierstra, and Martin Riedmiller.
\newblock Playing atari with deep reinforcement learning.
\newblock {\em arXiv preprint arXiv:1312.5602}, 2013.

\bibitem{mnih2015human}
Volodymyr {Mnih}, Koray {Kavukcuoglu}, David {Silver}, Andrei~A. {Rusu}, Joel
  {Veness}, Marc~G. {Bellemare}, Alex {Graves}, Martin {Riedmiller}, Andreas~K.
  {Fidjeland}, Georg {Ostrovski}, Stig {Petersen}, Charles {Beattie}, Amir
  {Sadik}, Ioannis {Antonoglou}, Helen {King}, Dharshan {Kumaran}, Daan
  {Wierstra}, Shane {Legg}, and Demis {Hassabis}.
\newblock Human-level control through deep reinforcement learning.
\newblock {\em Nature}, 518(7540):529--533, 2015.

\bibitem{mohsenianrad2010autonomous}
A.~Mohsenianrad, V.~W~S Wong, J.~Jatskevich, R.~Schober, and A.~Leongarcia.
\newblock Autonomous demand-side management based on game-theoretic energy
  consumption scheduling for the future smart grid.
\newblock {\em IEEE Transactions on Smart Grid}, 1(3):320--331, 2010.

\bibitem{mukerjee2015enabling}
M.~K. Mukerjee, J.~Hong, J.~Jiang, D.~Naylor, D.~Han, S.~Seshan, and H.~Zhang.
\newblock Enabling near real-time central control for live video delivery in
  cdns.
\newblock {\em ACM SIGCOMM Computer Communication Review}, 44(4):343--344,
  2015.

\bibitem{pan2009survey}
Sinno~Jialin Pan and Qiang Yang.
\newblock A survey on transfer learning.
\newblock {\em IEEE Transactions on knowledge and data engineering},
  22(10):1345--1359, 2009.

\bibitem{pang2017first}
Haitian Pang, Zhi Wang, Chen Yan, Qinghua Ding, and Lifeng Sun.
\newblock First mile in crowdsourced live streaming: A content harvest network
  approach.
\newblock In {\em Proceedings of the on Thematic Workshops of ACM Multimedia
  2017}, pages 101--109, 2017.

\bibitem{pang2018content}
Haitian Pang, Zhi Wang, Chen Yan, Qinghua Ding, Kun Yi, Jiangchuan Liu, and
  Lifeng Sun.
\newblock Content harvest network: Optimizing first mile for crowdsourced live
  streaming.
\newblock {\em IEEE Transactions on Circuits and Systems for Video Technology},
  29(7):2112--2125, 2018.

\bibitem{ran2018deepdecision}
Xukan Ran, Haolianz Chen, Xiaodan Zhu, Zhenming Liu, and Jiasi Chen.
\newblock Deepdecision: A mobile deep learning framework for edge video
  analytics.
\newblock In {\em IEEE INFOCOM 2018-IEEE Conference on Computer
  Communications}, pages 1421--1429. IEEE, 2018.

\bibitem{roman2018mobile}
Rodrigo Roman, Javier Lopez, and Masahiro Mambo.
\newblock Mobile edge computing, fog et al.: A survey and analysis of security
  threats and challenges.
\newblock {\em Future Generation Computer Systems}, 78:680--698, 2018.

\bibitem{savage1999end}
Stefan Savage, Andy Collins, Eric Hoffman, John Snell, and Thomas Anderson.
\newblock The end-to-end effects of internet path selection.
\newblock In {\em Proceedings of the conference on Applications, technologies,
  architectures, and protocols for computer communication}, pages 289--299,
  1999.

\bibitem{shi2016edge}
Weisong Shi, Jie Cao, Quan Zhang, Youhuizi Li, and Lanyu Xu.
\newblock Edge computing: Vision and challenges.
\newblock {\em IEEE internet of things journal}, 3(5):637--646, 2016.

\bibitem{silver2014deterministic}
David {Silver}, Guy {Lever}, Nicolas {Heess}, Thomas {Degris}, Daan {Wierstra},
  and Martin {Riedmiller}.
\newblock Deterministic policy gradient algorithms.
\newblock In {\em Proceedings of The 31st International Conference on Machine
  Learning}, pages 387--395, 2014.

\bibitem{VGG19}
Karen Simonyan and Andrew Zisserman.
\newblock Very deep convolutional networks for large-scale image recognition.
\newblock {\em arXiv preprint arXiv:1409.1556}, 2014.

\bibitem{song2014energy}
J.~Song, Y.~Cui, M.~Li, J.~Qiu, and R.~Buyya.
\newblock Energy-traffic tradeoff cooperative offloading for mobile cloud
  computing.
\newblock In {\em Quality of Service (IWQoS), 2014 IEEE 22nd International
  Symposium of}, pages 284--289. IEEE, 2014.

\bibitem{GoogleNet}
Christian Szegedy, Wei Liu, Yangqing Jia, Pierre Sermanet, Scott Reed, Dragomir
  Anguelov, Dumitru Erhan, Vincent Vanhoucke, and Andrew Rabinovich.
\newblock Going deeper with convolutions.
\newblock In {\em Proceedings of the IEEE conference on computer vision and
  pattern recognition}, pages 1--9, 2015.

\bibitem{InceptionNet}
Christian Szegedy, Vincent Vanhoucke, Sergey Ioffe, Jon Shlens, and Zbigniew
  Wojna.
\newblock Rethinking the inception architecture for computer vision.
\newblock In {\em Proceedings of the IEEE conference on computer vision and
  pattern recognition}, pages 2818--2826, 2016.

\bibitem{torfason2018towards}
Robert Torfason, Fabian Mentzer, Eirikur Agustsson, Michael Tschannen, Radu
  Timofte, and Luc Van~Gool.
\newblock Towards image understanding from deep compression without decoding.
\newblock {\em arXiv preprint arXiv:1803.06131}, 2018.

\bibitem{vermorel2005multi}
Joannes Vermorel and Mehryar Mohri.
\newblock Multi-armed bandit algorithms and empirical evaluation.
\newblock In {\em European conference on machine learning}, pages 437--448.
  Springer, 2005.

\bibitem{wang2015understanding}
H.~Wang, F.~Xu, Y.~Li, P.~Zhang, and D.~Jin.
\newblock Understanding mobile traffic patterns of large scale cellular towers
  in urban environment.
\newblock In {\em Proceedings of the 2015 ACM Conference on Internet
  Measurement Conference}, pages 225--238. ACM, 2015.

\bibitem{Wang2017Towards}
X.~Wang, X.~Chen, and W.~Wu.
\newblock Towards truthful auction mechanisms for task assignment in mobile
  device clouds.
\newblock In {\em IEEE INFOCOM 2017 - IEEE Conference on Computer
  Communications}, pages 1--9, 2017.

\bibitem{wang2016propagation}
Zhi Wang, Lifeng Sun, Miao Zhang, Haitian Pang, Erfang Tian, and Wenwu Zhu.
\newblock Propagation-and mobility-aware d2d social content replication.
\newblock {\em IEEE Transactions on Mobile Computing}, 16(4):1107--1120, 2016.

\bibitem{xu2015forecasting}
J.~Xu, M.~van~der Schaar, J.~Liu, and H.~Li.
\newblock Forecasting popularity of videos using social media.
\newblock {\em Selected Topics in Signal Processing, IEEE Journal of},
  9(2):330--343, 2015.

\bibitem{ye2021infocom}
Jiahui Ye, Zichun Li, Zhi Wang, Zhuobin Zheng, Han Hu, and Wenwu Zhu.
\newblock Joint cache size scaling and replacement adaptation for small content
  providers.
\newblock In {\em Proceedings of IEEE INFOCOM}, 2021.

\bibitem{yuan2019adversarial}
Xiaoyong Yuan, Pan He, Qile Zhu, and Xiaolin Li.
\newblock Adversarial examples: Attacks and defenses for deep learning.
\newblock {\em IEEE transactions on neural networks and learning systems},
  2019.

\bibitem{zhang2015crowdsourced}
Cong Zhang and Jiangchuan Liu.
\newblock On crowdsourced interactive live streaming: a twitch. tv-based
  measurement study.
\newblock In {\em Proceedings of the 25th ACM Workshop on Network and Operating
  Systems Support for Digital Audio and Video}, pages 55--60, 2015.

\bibitem{zhangunderstand}
Lei Zhang, Feng Wang, and Jiangchuan Liu.
\newblock Understand instant video clip sharing on mobile platforms: Twitter's
  vine as a case study.
\newblock In {\em ACM Network and Operating System Support on Digital Audio and
  Video Workshop (NOSSDAV)}, 2014.

\bibitem{zhu2011multimedia}
Wenwu Zhu, Chong Luo, Jianfeng Wang, and Shipeng Li.
\newblock Multimedia cloud computing.
\newblock {\em IEEE Signal Processing Magazine}, 28(3):59--69, 2011.

\bibitem{zwolenski2014digital}
Matt Zwolenski, Lee Weatherill, et~al.
\newblock The digital universe: Rich data and the increasing value of the
  internet of things.
\newblock {\em Journal of Telecommunications and the Digital Economy}, 2(3):47,
  2014.

\end{thebibliography}

\end{document}